\documentclass[%
 reprint,
 superscriptaddress,
 amsmath,amssymb,
 aps,
pre,
 floatfix
]{revtex4-2}

\usepackage{graphicx}
\usepackage{float}
\usepackage{dcolumn}
\usepackage{bm}
\usepackage{booktabs}
\usepackage{hyperref}
\hypersetup{
    colorlinks=true, 
    linktoc=all,     
    linkcolor=black,  
    allcolors=black,
}
\usepackage{array}
\newcolumntype{P}[1]{>{\centering\arraybackslash}p{#1}}

\usepackage{ulem}
\usepackage{xcolor}
\definecolor{forestgreen}{rgb}{0.33,0.61,0.34}

\begin{document}

\preprint{APS/123-QED}

\title{Generative models of simultaneously heavy-tailed distributions \\ of interevent times on nodes and edges}

\author{Elohim Fonseca dos Reis}
\affiliation{Department of Mathematics, State University of New York at Buffalo, Buffalo, New York, USA}
\author{Aming Li}
\affiliation{Department of Zoology, University of Oxford, Oxford, UK}
\affiliation{Department of Biochemistry, University of Oxford, Oxford, UK}
\author{Naoki Masuda}
\email{naokimas@buffalo.edu}
\affiliation{Department of Mathematics, State University of New York at Buffalo, Buffalo, New York, USA}
\affiliation{Computational and Data-Enabled Science and Engineering Program, State University of New York at Buffalo, Buffalo, New York, USA}
\affiliation{Faculty of Science and Engineering, Waseda University, Tokyo, Japan}

\date{\today}

\begin{abstract}
Intervals between discrete events representing human activities, as well as other types of events, often obey heavy-tailed distributions, and their impacts on collective dynamics on networks such as contagion processes have been intensively studied.
The literature supports that such heavy-tailed distributions are present for interevent times associated with both individual nodes and individual edges in networks.
However, the simultaneous presence of heavy-tailed distributions of interevent times for nodes and edges is a non-trivial phenomenon, and its origin has been elusive.
In the present study, we propose a generative model and its variants to explain this phenomenon.
We assume that each node independently transits between a high-activity and low-activity state according to a continuous-time two-state Markov process and that, for the main model, events on an edge occur at a high rate if and only if both end nodes of the edge are in the high-activity state.
In other words, two nodes interact frequently only when both nodes prefer to interact with others.
The model produces distributions of interevent times for both individual nodes and edges that resemble heavy-tailed distributions across some scales.
It also produces positive correlation in consecutive interevent times, which is another stylized observation for empirical data of human activity.
We expect that our modeling framework provides a useful benchmark for investigating dynamics on temporal networks driven by non-Poissonian event sequences.
\end{abstract}

\maketitle


\section{\label{sec:introduction}Introduction}

Dynamics contacts as well as the static structure of social contact networks govern how humans or animals gather, communicate, and act.
Many techniques from temporal networks have been proven useful for describing and utilizing data of time-varying networks \cite{HolmeSaramaki2012PhysRep, Holme2015EPJB, Masuda2016Book, Karsai2018book, HolmeSaramaki2019book}.
The time between two consecutive contacts, called the interevent time (IET), is a key quantity to characterize temporal networks and dynamics on them.
Myriad human activities, such as online chats, email correspondence, mobility, web browsing, and broker trading, have heavy-tailed distributions of IETs \cite{Barabasi2005Nature, Vazquez2006PRE, HolmeSaramaki2012PhysRep, Karsai2018book}.
This observation implies that sequences of discrete events that an individual node or edge in a network experiences obeys non-Poissonian statistics.
By contrast, in most cases, stochastic processes on static networks implicitly assume that events such as infection or broadcasting occur according to Poisson processes, with which IETs obey an exponential distribution.
Therefore, the non-Poissonian nature of event sequences in empirical data inevitably urges us to reconsider our understanding of stochastic dynamical processes on networks.
In fact, effects of heavy-tailed distributions of IETs on epidemic processes \cite{Min2011PRE, KarsaiPRE2011, Rocha2011PlosComputBio, Miritello2011PRE, Masuda2013F1000prime, Jo2014PRX, PastorSatorras2015RMF, Masuda2017book}, opinion dynamics \cite{Wu2010PhysicaA, Takaguchi2011PRE, FernandezGracia2011PRE, Nishi2014EPL}, evolutionary game dynamics \cite{Li2020NatComm}, and cascade processes \cite{Karimi2013PhysicaA, Takaguchi2013PLoSOne, Backlund2014PRE, Unicomb2020arXiv}, random walks \cite{Hoffmann2012PRE, Starnini2012PRE, Speidel2015PRE, Masuda2017PhysRep}, to name a few, have been studied.
There are also a number of generative mechanisms and descriptive models for heavy-tailed distributions of IETs including priority queuing models \cite{Barabasi2005Nature, Vazquez2005PRL, Vazquez2006PRE, Grinstein2006PRL, Grinstein2008PRE, Masuda2009PRE, Oliveira2009PhysicaA, Jo2012PRE}, mixture of exponentials \cite{MasudaHolme2020PRR, Okada2020RSOS, Jiang2016JStatMech}, those supplied by circadian and weekly rhythms \cite{Malmgren2008PNAS}, and other self-exciting processes \cite{Malmgren2009Science, Masuda2013bookchap}.

\begin{figure*}[ht]
\includegraphics{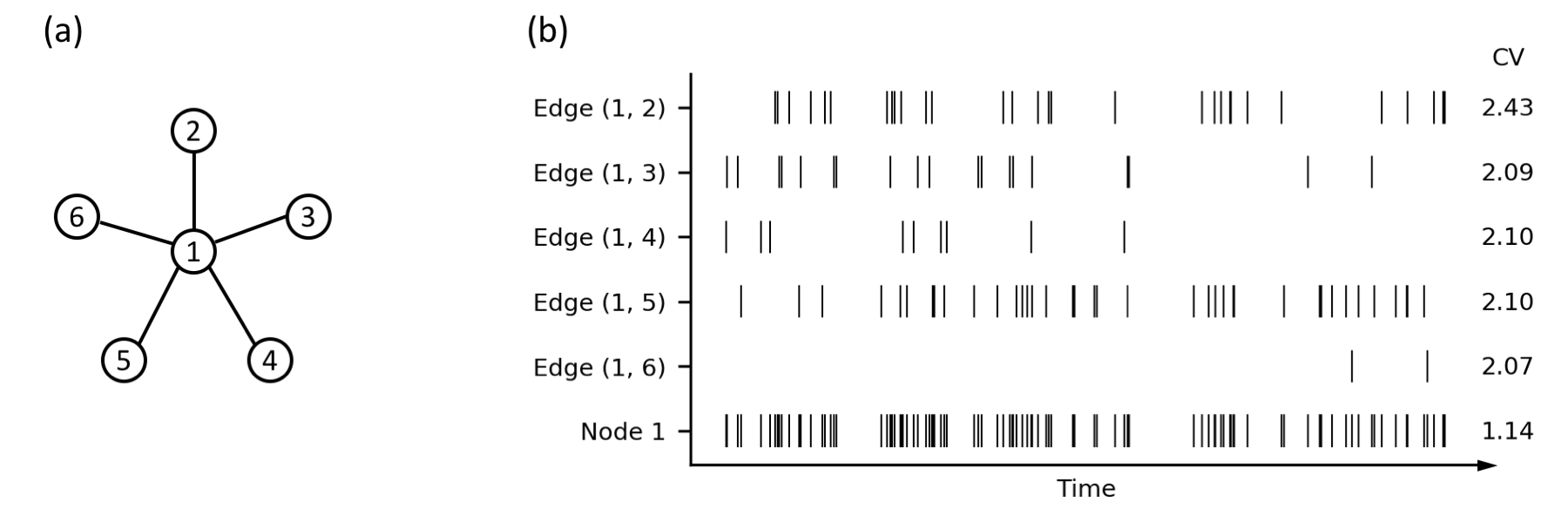}
\caption{\label{fig:powerlaw_schematic}
	Schematic illustration of the superposition of event sequences on edges.
	(a) Star network of six nodes.
	(b) Sequence of events generated by a power-law distribution of IET on each edge of the star network and the sequence of events on node $1$ in (a).
	The CV of IETs, which we calculate from the first approximately $10^6$ events on each edge, is shown in (b) for the five edges and node 1.
	We used the power-law distribution of IETs for edges given by $p(\tau) = (\alpha - 1)/(1+\tau)^{\alpha}$, where $\alpha=3.5$.
	According to an equilibrium renewal process \cite{Cox1962book, Masuda2016Book}, we draw the time to the first event on each edge from the distribution of waiting times given by $p^\text{w}(t) = (\alpha - 2)/(1+t)^{\alpha-1}$.
	}
\end{figure*}

Heavy-tailed distributions of IETs are commonly found for single nodes \cite{Barabasi2005Nature, Vazquez2006PRE, Malmgren2008PNAS, Karsai2012SciRep, Jiang2016JStatMech, Masuda2011PRX, Eckmann2004PNAS} and single edges \cite{Karsai2012PlosOne, Saramaki2015EPJB_2, KarsaiPRE2011, Karsai2018book, HolmeSaramaki2012PhysRep}.
Such a distribution for a node implies that the sequence of event times for the node, regardless of the identity of the neighbor, obeys a non-Poissonian, heavy-tailed statistics.
In fact, the distribution of IETs for both nodes and edges in a single data set are often heavy-tailed (see Section \ref{sec:data_analysis} for examples).
However, the presence of heavy-tailed IET distributions for both edges and nodes in the same network is not trivial.
Consider a node, denoted by $v$, that interacts with its $k$ neighbors, and assume that the sequence of IETs on each edge incident to $v$ independently obeys a heavy-tailed distribution.
The superposition of the $k$ event sequences yields the sequence of $v$'s events.
This situation is illustrated in Fig.~\ref{fig:powerlaw_schematic}.
In Fig.~\ref{fig:powerlaw_schematic}(a), node $v$ has $k=5$ neighbors.
We have produced the sequence of events on each of the five edges assuming a power-law distribution of IETs, as shown in Fig.~\ref{fig:powerlaw_schematic}(b).
The sequence of events shown in the bottom of Fig.~\ref{fig:powerlaw_schematic}(b) is that for $v$, which one obtains by superposing the sequences of events on the $k$ edges.
In general, the distribution of IETs for $v$ is less heavy-tailed than that for a single edge.
This is because the superposition of independent point processes (more precisely, renewal processes) obeying heavy-tailed distributions roughly approaches, albeit not precisely, a Poisson process as one increases $k$ \cite{Lindner2006PRE, Cateau2006PRL}.

Recently, we proposed a model that generates heavy-tailed distributions of IETs for both edges and nodes \cite{Hiraoka2020PRR}.
The model assumes that each node is activated at discrete times according to a renewal process that draws the inter-activation time from a power-law distribution.
Then, at each time step, activated nodes are uniformly randomly selected to communicate with simultaneously activated neighbors.
By construction, this model produces a heavy-tailed distribution of IETs for individual nodes.
Although it is less trivial, IETs for edges also obey approximately heavy-tailed distributions.  
However, this model does not explain why we find heavy-tailed distributions of IETs for both nodes and edges in empirical data.
Realistic mechanisms of the simultaneous presence of heavy-tailed IET distributions for nodes and edges are underexplored \cite{Karsai2018book}.

In the present study, we propose a model of time-stamped event sequences on networks that is based on a latent state dynamics of nodes.
In our model, each node switches between two states called the high-activity and low-activity states according to a Markov process in continuous time.
We assume that if both nodes are in the high-activity state, events occur according to a Poisson process at a higher rate, otherwise at a lower rate.
The rationale behind the model is that events between two individuals may happen more frequently when both individuals are motivated to interact with others than otherwise.
We show that our model produces distributions of IETs with large dispersions for both single nodes and edges, resembling empirical data.

\section{\label{sec:data_analysis}Simultaneously large variability of interevent times on nodes and edges is common and non-trivial} 

Before presenting and analyzing our model, in this section we provide empirical evidence that heavy-tailed distributions of IETs are simultaneously present for edges and nodes in single data sets.
We use data collected by the SocioPatterns collaboration \cite{Genois2018EPJ}.
The survival function of the IETs (i.e., probability that the IET, $\tau$, is larger than the specified value) for individual edges and nodes for social contact data in a primary school \cite{Genois2018EPJ}, which we refer to as PrimarySchool, is shown in Fig.~\ref{fig:survival}(a), and \ref{fig:survival}(b), respectively. 
(See Appendix \ref{appendix:data_survival} for the results for the other data sets.)
The relatively slow decay in Fig.~\ref{fig:survival} suggests heavy-tailed distributions for both edges and nodes across some scales of $\tau$.

\begin{figure*}[ht]
	\includegraphics{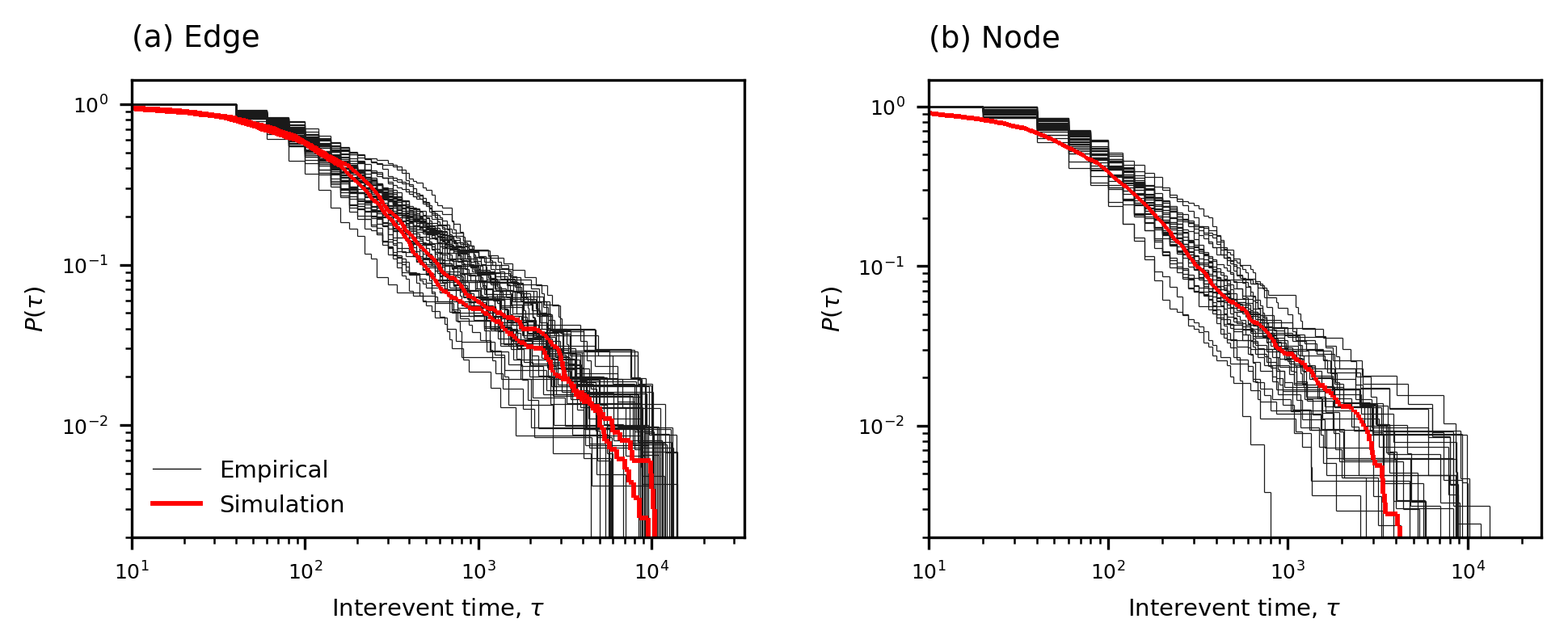}
	\caption{\label{fig:survival}
		Survival function, $P(\tau)$, of IETs on (a) edges and (b) nodes for the PrimarySchool data set (thin lines) and simulation of our model (thick lines).
		In this and the following analyses of the empirical data, we treated the data with two steps.
		First, we aggregated consecutive events with 20 s duration separately for each edge (a) or node (b); the temporal resolution of the original data set is 20 s, i.e., the social contacts are measured every 20 s.
		We perform this first step because we are analyzing the IETs but not the duration of events.
		For instance, we aggregated a sequence of event times \{20, 40, 60, 100, 200, 220\} (in s) between two given nodes into a sequence with three contact events as \{20, 100, 200\}.
		In other words, the first event on this edge occurs at $t=20$ s and lasts for 60 s, the second event at $t=100$ s lasts for 20 s, and the third event at $t=200$ s lasts for 40 s.
		Second, to circumvent the effects of the circadian rhythm, we removed IETs larger than eight hours.
		In both (a) and (b), we only considered edges that had at least 100 events, after aggregating consecutive events with 20 s duration and removing IETs larger than 8 h.
		}
\end{figure*}

We quantified the dispersion of IET distributions by the coefficient of variation (CV).
The CV of a distribution is defined as the standard deviation divided by the mean.
For an IET distribution, one can write
\begin{eqnarray}
	\mathrm{CV} = \sqrt{\frac{\langle \tau^2 \rangle}{\langle \tau \rangle^2} - 1},
\label{eq:CV}
\end{eqnarray}
where $\langle \cdot \rangle$ represents the average over edges or nodes.
A Poisson process produces an exponential IET distribution, which yields $\mathrm{CV} = 1$.
A periodic process yields $\mathrm{CV} = 0$.
A heavy-tailed distribution yields a large value of CV.
Table \ref{tab:cv} shows the mean and standard deviation of the CV of the IET distribution for edges and nodes for each data set.
Table \ref{tab:cv} indicates that all data sets yield CV values considerably larger than 1 for both edges and nodes.
Therefore, the simultaneous presence of heterogeneous distributions (i.e., with a larger dispersion than the Poisson case) of both edge's and node's IETs seems to be common.

\begin{table*}[t]
\caption{\label{tab:cv}
	CV of IETs on edges and nodes for SocioPatterns data sets.
	``Node CV, shuffled'' corresponds to the node's CV calculated after the timeline interevent shuffling.
	The data originate from a primary school (PrimarySchool) \cite{Gemmetto2014BMC, Stehle2011PlosOne}, a scientific conference (SFHH) \cite{Genois2018EPJ}, a workplace (Office15) \cite{Genois2018EPJ}, a hospital (Hospital) \cite{Vanhems2014PlosOne}, and a high school in two different years (HighSchool12 \cite{Fournet2014PlosOne} and HighSchool13 \cite{Mastrandrea2015PlosOne}).	
	The CV values shown are the average $\pm$ standard deviation.
	In this table and the following figures and tables using the same data sets, we used the edges that have at least 100 events, after aggregating consecutive events with 20 s duration and removing IETs larger than 8 h; we used nodes that have, among its incident edges, at least one edge with at least 100 events and all the other edges with at least 10 events.
	}
\begin{ruledtabular}
   \begin{tabular}{cP{2.2cm}P{2.2cm}P{2.2cm}P{2.2cm}P{2.2cm}P{2.2cm}}
           & PrimarySchool & SFHH           & Office15      & Hospital      & HighSchool12  & HighSchool13   \\
   \cmidrule{1-7}
   Edge CV, original     & $2.7 \pm 0.6$ & $2.0 \pm 0.9$  & $2.5 \pm 0.8$ & $1.6 \pm 0.4$ & $2.7 \pm 0.6$ & $2.5 \pm 0.6$  \\
   Node CV, original     & $3.2 \pm 1.3$ & $2.2 \pm 1.2$  & $2.8 \pm 0.8$ & $1.9 \pm 0.9$ & $2.8 \pm 0.6$ & $2.6 \pm 0.6$  \\
   \cmidrule{1-7}
   Node CV, shuffled & $1.7 \pm 0.7$ & $1.8 \pm 1.1$  & $2.8 \pm 0.8$ & $1.5 \pm 0.6$ & $2.7 \pm 0.6$ & $2.3 \pm 0.5$  \\
   $\Delta$ (i.e., relative deviation)  & $-46$\%       & $-17$\%        & $-3$\%        & $-20$\%       & $-5$\%        & $-11$\%        \\
   \end{tabular}
\end{ruledtabular}
\end{table*}

We hypothesize that the correlation of the IET between different edges sharing a node contributes to large values of the CV for individual nodes.
Therefore, for each data set, we uniformly randomly shuffled the IETs on each edge within each day of recording, preserving the time of the first and last events of each day on the edge.
This shuffling method is equivalent to the timeline interevent shuffling in an instant-event temporal network (formally named $\mathrm{P}[\bm{\pi}_{\mathcal{L}}(\bm{\Delta \tau}), \bm{t}^1]$) described in Ref.~\cite{Gauvin2018arXiv}.
Because this shuffling procedure preserves the distribution of IETs on each edge, the edge's CV is unchanged.
However, it affects the IETs and hence the CV for individual nodes.
The node's CV calculated from the shuffled data and the relative deviation defined by $\Delta = 100\% \times (\text{shuffled}-\text{original})/\text{original}$, are shown in Table \ref{tab:cv}.
For all data sets, the CV for the nodes consistently decreases when one shuffles the IETs in the majority of the data sets, although the differences are statistically insignificant.
Therefore, the shuffling tends to destroy the heavy-tailed nature of the IET sequences for individual nodes.

To further support that the simultaneous presence of heavy-tailed distributions of IETs on edges and nodes is nontrivial, we assess a common approach to generate event sequences on each edge according to an independent renewal process with a power-law distribution, $p(\tau)$, of IETs.
We call this model the stochastic temporal network model \cite{Masuda2016Book}.
Consider the power-law distribution of IETs given by
\begin{eqnarray}
	 p(\tau) = \frac{\alpha - 1}{(1+\tau)^{\alpha}},
\label{eq:powerlaw}
\end{eqnarray}
where $\alpha$ is a parameter.
The first and second moments of $p(\tau)$ are given by $\langle \tau \rangle = 1/(\alpha-2)$, where $\alpha >2$, and $\langle \tau^2 \rangle = 2/(\alpha - 3)(\alpha - 2)$, where $\alpha >3$, respectively.
Therefore, the CV of IETs is given by
\begin{eqnarray}
	\text{CV} = \sqrt{ \frac{\alpha - 1}{\alpha - 3} },
\label{eq:powerlaw_CV}
\end{eqnarray}
where $\alpha>3$.
We ran simulations with $\alpha=3$ and $\alpha=3.5$ for a node with $k=2$, $k=5$, and $k=10$ neighbors, and calculated the CV for IETs on the node for each combination of $\alpha$ and $k$.
Equation \eqref{eq:powerlaw_CV} predicts an edge's CV equal to 2.24 when $\alpha=3.5$ and its divergence for $\alpha \leq 3$, which is consistent with the numerical results shown in Table \ref{tab:power-law}. 
By contrast, the CV for the node is considerably smaller than that for edge and decreases towards 1 as $k$ increases.
Therefore, the stochastic temporal network model does not produce simultaneously heavy-tailed distributions of IETs for edges and nodes.

\begin{table}[ht]
   \caption{\label{tab:power-law}
	CV of IETs on edges and nodes obtained from the stochastic temporal network model.
	We assume an equilibrium renewal process, so we draw the time to the first event on each edge from the distribution of waiting times, as in Fig.~\ref{fig:powerlaw_schematic}.
	We calculated the mean and standard deviation of the CVs on the basis of 100 realizations of the simulation.
	We stop each realization when all edges have obtained at least $10^6$ events.
	}
   \begin{ruledtabular}
   \begin{tabular}{cccc}
      $\alpha$ & $k$  &  Edge CV              &  Node CV              \\ 
   \cmidrule{1-4}
               &  2   & $ 3.9 \pm 1.3 $       & $ 1.7 \pm 0.0 $       \\
      3        &  5   & $ 3.8 \pm 0.6 $       & $ 1.2 \pm 0.0 $       \\
               &  10  & $ 3.9 \pm 0.6 $       & $ 1.1 \pm 0.0 $       \\
   \cmidrule{1-4}
               &  2   & $ 2.2 \pm 0.2 $       & $ 1.4 \pm 0.0 $       \\
      3.5      &  5   & $ 2.2 \pm 0.1 $       & $ 1.1 \pm 0.0 $       \\
               &  10  & $ 2.2 \pm 0.1 $       & $ 1.1 \pm 0.0 $       \\
   \end{tabular}
   \end{ruledtabular}
\end{table}

\section{\label{sec:model}Model} 

We propose a model of node behavior that aims to simultaneously produce large CVs for IETs on individual edges and nodes.
Consider a static network.
The model is based on two main assumptions.
First, we assume that each node stochastically switches between two states, called the high-activity state (denoted by $h$) and the low-activity state (denoted by $\ell$).
The plausibility of this assumption is supported by various empirical and modeling studies \cite{Malmgren2008PNAS, Malmgren2009Science, Karsai2012SciRep, Raghavan2014IEEETransComputSocSys, Jiang2016JStatMech, Clementi2008ProcACM, Clementi2010SIAM, Okada2020RSOS, MasudaHolme2020PRR}.
Second, we assume that two nodes adjacent by an edge in the static network have a contact event much more likely when both nodes are in the high-activity state than otherwise.
The intuition behind this assumption is that a pairwise human contact event may be much more likely to occur when both individuals are motivated to interact than otherwise.

Each node switches between $h$ and $\ell$ according to a two-state continuous-time Markov process.
In other words, the node switches to the opposite state according to a Poisson process whose rate depends on the current state.
We denote by $r_{h \rightarrow \ell}$ the rate at which a node in state $h$ changes to state $\ell$, and similar for $r_{\ell \rightarrow h}$.
We assume that different nodes share the same $r_{h \rightarrow \ell}$ and $r_{\ell \rightarrow h}$ values but are associated with independent two-state Markovian processes.

\begin{figure}[ht]
	\includegraphics{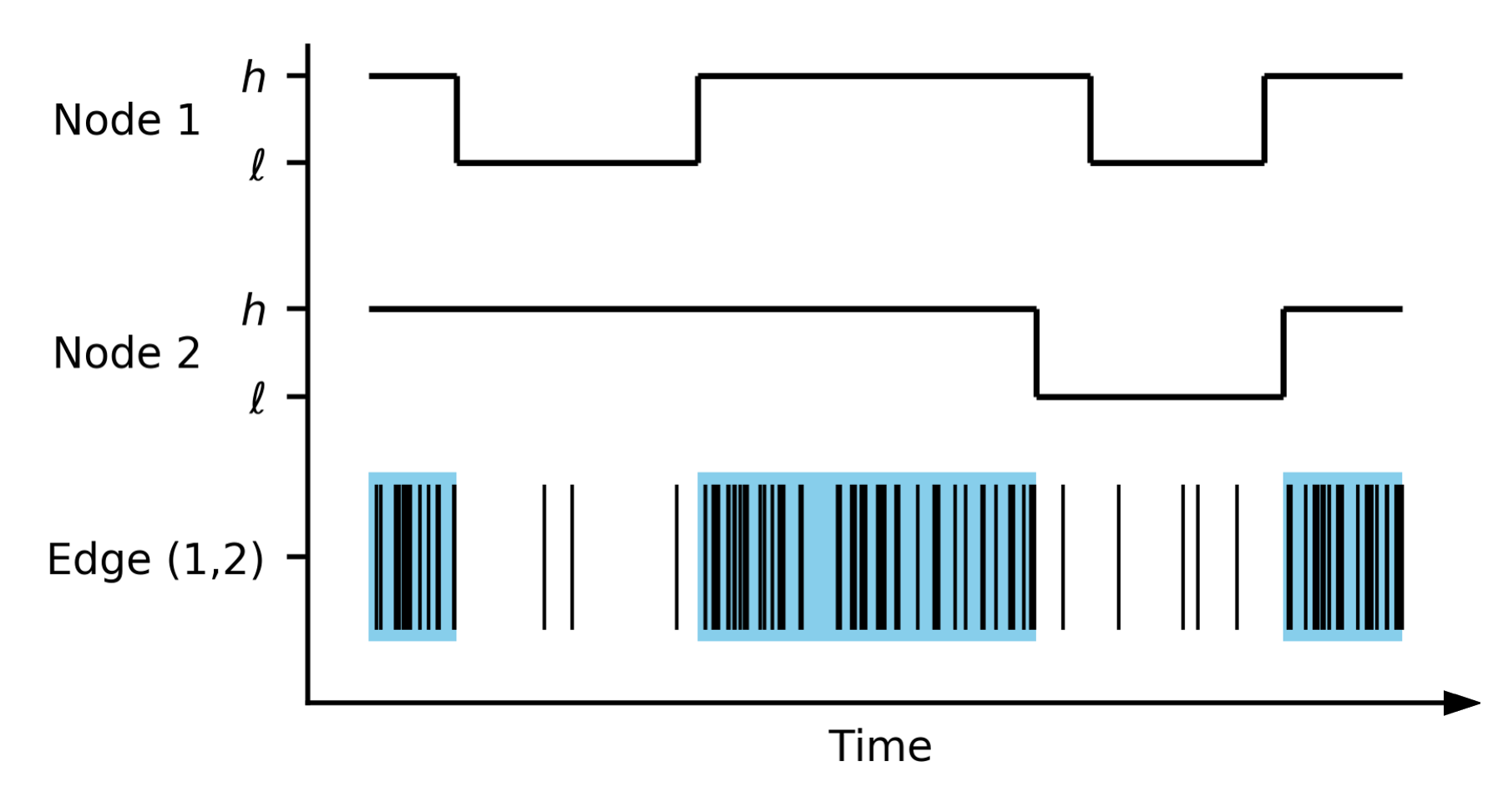}
	\caption{\label{fig:edge_dynamics}
		Schematic illustration of the model.
		The events between two nodes occur at a higher rate $\lambda_h$ if and only if both are in the high-activity state (corresponding to the shaded time windows).
		Otherwise, events occur at a lower rate $\lambda_{\ell}$ (corresponding to the unshaded time windows).
	}
\end{figure}

If both adjacent nodes are in the $h$ state, then events on the edge are assumed to occur according to a Poisson process at a higher rate denoted by $\lambda_h$.
Otherwise, the edge produces events according to a Poisson process at a lower rate denoted by $\lambda_{\ell}$ ($< \lambda_h$).
This process is schematically shown in Fig.~\ref{fig:edge_dynamics}.

The master equation for the probability that a node is in state $h$, denoted by $p_h(t)$, where $t$ represents time, is given by
\begin{eqnarray}
	\frac{\mathrm{d} p_h(t)}{\mathrm{d} t} = r_{\ell \rightarrow h} [1-p_h(t)] - r_{h \rightarrow \ell} \, p_h(t).
\label{eq:master_equation}
\end{eqnarray}
Therefore, the stationary probability of finding a node in state $h$ is given by
\begin{eqnarray}
	p_h^* = \frac{r_{\ell \rightarrow h}}{r_{h \rightarrow \ell} + r_{\ell \rightarrow h}}.
\label{eq:p_h}
\end{eqnarray}
The stationary probability of finding a node in state $\ell$ is $p_{\ell}^*=1-p_h^*$.

\section{Results} \label{sec:results}

\subsection{\label{sec:results_orig}The model produces simultaneously large variability of IETs on edges and nodes}

We numerically simulated the model to generate sequences of IETs and computed the survival function and the CV of IETs for individual edges and nodes.
Apart from the structure of the static network, our model has four parameters, $r_{h \rightarrow \ell},\, r_{\ell \rightarrow h},\, \lambda_h, \,\text{and}\, \lambda_{\ell}$.
Equation \eqref{eq:p_h} yields $r_{\ell \rightarrow h} = r_{h \rightarrow \ell} \, p_h^*/(1 - p_h^*)$.
We write $\lambda_{\ell}$ in terms of $\lambda_h$ as
\begin{eqnarray}
	\lambda_{\ell} = \gamma \lambda_h,
\end{eqnarray}
where $0<\gamma<1$.
Simultaneously changing ($r_{h \rightarrow \ell},\, r_{\ell \rightarrow h},\, \lambda_h,\, \lambda_{\ell}$) to ($c r_{h \rightarrow \ell},\, c r_{\ell \rightarrow h},\, c \lambda_h,\, c \lambda_{\ell}$), where $c > 0$, is equivalent to not changing these four parameters and changing the time from $t$ to $ct$.
Therefore, without loss of generality, we set $\lambda_h = 1$, unless we state otherwise.
In the following simulations, for fixed values of $r_{h \rightarrow \ell}$, we varied $\gamma$ and $p_h^*$.
Using the Gillespie algorithm \cite{Gillespie1977JPhysChem}, we generated events on the edges until all edges had at least $10^6$ events.

We used a star network composed of a node $v$ and its $k$ neighbors.
Initially, each of the $k+1$ nodes is independently in state $h$ or $\ell$ with probability $p_h^*$ or $(1-p_h^*)$, respectively.
Then, we run a continuous-time Markov process independently for each node.
At each time, depending on the state of each edge, we generate events on the edge at rate $\lambda_h$ or $\lambda_{\ell}$.

It should be noted that a next event on an edge may not be simply produced as a single Poisson process.
For example, suppose that the two nodes connected by an edge are both in state $h$.
Then, the time to the next event on this edge obeys the exponential distribution $p(\tau)= \lambda_h e^{- \lambda_h \tau}$.
However, if either node switches to state $\ell$ before the next event occurs, then the distribution $p(\tau)= \lambda_h e^{- \lambda_h \tau}$ is no longer relevant, and the time to the next event now obeys $p(\tau) = \lambda_{\ell} e^{-\lambda_{\ell} \tau}$.
Note that the point process producing the events is Poissonian and therefore memoryless if conditioned on the nodes' states.
Therefore, if either node has switched to state $\ell$ at time $\tilde{t}$, then the time to the next event that obeys $p(\tau) = \lambda_{\ell} e^{-\lambda_{\ell} \tau}$ is added to $\tilde{t}$ to set the next event time.
The same applies to the case in which one node is in state $h$ and the other node is in state $\ell$ (therefore, the time to the next event obeys $p(\tau)=\lambda_{\ell} e^{-\lambda_{\ell} \tau}$), and then the latter node switches to state $h$ such that the time to the next event now obeys $p(\tau)=\lambda_h e^{-\lambda_h \tau}$.

We consider a star network composed of a node $v$ and its two neighbors as an example.
We set $r_{h \rightarrow \ell}=2 \times 10^{-5}$, $r_{\ell \rightarrow h} = 4.7 \times 10^{-5}$, $\lambda_h = 6 \times 10^{-3}$, and $ \lambda_{\ell} = 3.5 \times 10^{-4}$, which yields $p_h^* \approx 0.7$ and $\gamma \approx 0.06$.
The survival function of IETs produced by the model is shown by the thick lines for the two edges and node $v$ in Fig.~\ref{fig:survival}(a) and Fig.~\ref{fig:survival}(b), respectively.
The survival function for both the two edges and $v$ decays more slowly than exponentially, roughly consistent with the non-Poissonian behavior observed in the empirical data (thin lines in Fig.~\ref{fig:survival}).
For our model, the CV for the two edges is equal to 2.8 and 3.0, and that for $v$ is equal to 2.6.
These values of CV are statistically within the ranges of the CV for the PrimarySchool data set (see Table \ref{tab:cv}).

\begin{figure*}[ht]
	\includegraphics{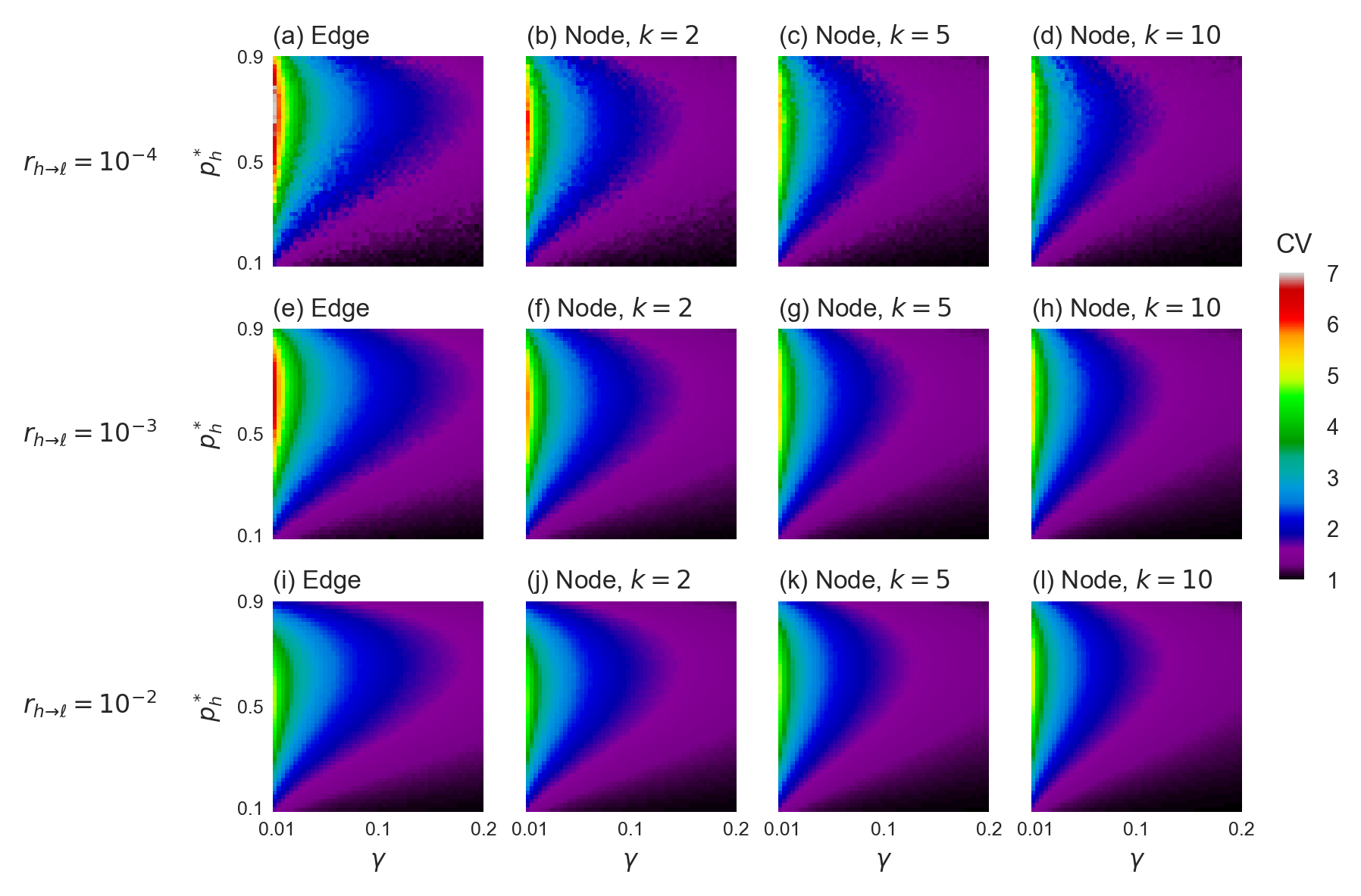}
	\caption{\label{fig:CVheatmap_orig}
	CV values for IETs generated by our original model.
	We set $r_{h \rightarrow \ell} = 10^{-4}$ in panels (a), (b), (c), and (d), $r_{h \rightarrow \ell} = 10^{-3}$ in panels (e), (f), (g), and (h), and $r_{h \rightarrow \ell} = 10^{-2}$ in panels (i), (j), (k), and ($\ell$).
	We simulated an edge (panels (a), (e), and (i)), a node with $k=2$ neighbors (panels (b), (f), and (j)), a node with $k=5$ neighbors (panels (c), (g), and (k)), and a node with $k=10$ neighbors (panels (d), (h), and ($\ell$)).
	For each set of parameter values, we generated IETs on the edges until all edges had at least $10^6$ events.
	}
\end{figure*}

To examine different parameter values of the model, we varied  $\gamma$ and $p_h^*$ for each of three values of $r_{h \rightarrow \ell}$ (i.e., $r_{h \rightarrow \ell} = 10^{-4}$, $r_{h \rightarrow \ell} = 10^{-3}$, and $r_{h \rightarrow \ell} = 10^{-2}$) and three values of $k$ (i.e., $k=2$, $k=5$, $k=10$).
We show the CV values in Fig.~\ref{fig:CVheatmap_orig}.
The figure indicates that the model produces large CV values simultaneously for edges and nodes in a broad parameter region.
In particular, the CV is large when $\gamma$ ($= \lambda_{\ell}/\lambda_h$) is small and $p_h^*$ is near 0.7, across the range of $r_{h \rightarrow \ell}$ and $k$.

\subsection{\label{sec:theoretical_cv}Analytical evaluation of the CV of interevent times} 

In this section, we provide an analytical account for the CV values observed in section \ref{sec:results_orig}.
The state of the edge is specified by the states of two nodes forming the edge.
We denote by $h^2$ when both nodes are in state $h$, and likewise for $h\ell$ and $\ell^2$.
The edge state obeys a three-state Markov process in the state space $\mathcal{S}_1 = \{h^2, h\ell, \ell^2 \}$, where we do not distinguish between $h\ell$ and $\ell h$, and represent both by $h\ell$.
In our model, the IET distribution conditioned on the edge's state is given by $g(\tau|h^2) = \lambda_h e^{-\lambda_h \tau}$ and $g(\tau|\ell^2) = g(\tau|h\ell) = \lambda_{\ell} e^{-\lambda_{\ell} \tau}$, where $g(\tau | \cdot)$ represents the distribution of IETs conditioned on the edge's state.

The Markov process of node activity is independent for different nodes.
Therefore, two nodes are simultaneously in state $h$ in the equilibrium with probability $p_h^{*2}$.
The mean number of events produced when both nodes are in state $h$ is given by $\lambda_h p_h^{*2} T$, where $T$ is the observation time.
The mean number of events produced when either node is in state $\ell$ is given by $\lambda_{\ell} (1-p_h^{*2}) T$.
Therefore, an IET is produced at rate $\lambda_h$ and $\lambda_{\ell}$ with probability $\lambda_h p_h^{*2} / \Omega_1$ and $\lambda_{\ell} (1-p_h^{*2}) / \Omega_1$, respectively, where $\Omega_1 = \lambda_h p_h^{*2} + \lambda_{\ell} (1-p_h^{*2})$.
By combining these contributions and ignoring IETs during which the edge's state changes, we obtain the probability density function (PDF) of IETs for an edge as a mixture of two exponential distributions as
\begin{eqnarray} \label{eq:edge_PDF}
	f_{\text{edge}}(\tau) = \frac{\lambda_h p_h^{*2}}{\Omega_1} \lambda_h e^{-\lambda_h \tau} + \frac{\lambda_{\ell} (1-p_h^{*2})}{\Omega_1} \lambda_{\ell} e^{-\lambda_{\ell} \tau}.
\end{eqnarray}
The first two moments of this PDF are given by
\begin{eqnarray} \label{eq:edge_first_moment}
	\langle \tau \rangle_{\text{edge}} \equiv \int_0^\infty \tau f_{\text{edge}} (\tau) d \tau = \frac{\,1\,}{\,\,\Omega_1}.
\end{eqnarray}
and
\begin{equation} \label{eq:edge_second_moment}
	\langle \tau^2 \rangle_{\text{edge}} \equiv \int_0^\infty \tau^2 f_{\text{edge}}(\tau) d \tau  = \frac{\,2\,}{\,\,\Omega_1}\left[ \frac{p_h^{*2}}{\lambda_h} + \frac{(1-p_h^{*2})}{\lambda_{\ell}} \right].
\end{equation}
By substituting Eqs.~\eqref{eq:edge_first_moment} and \eqref{eq:edge_second_moment} into Eq.~\eqref{eq:CV}, we obtain
\begin{eqnarray} \label{eq:edge_CV}
	\mathrm{CV}_{\text{edge}} = \sqrt{ 1 + \frac{2 p_h^{*2} (1-p_h^{*2}) (1-\gamma)^2}{\gamma} },
\end{eqnarray}
where $\rm CV_{\rm edge}$ is the CV for the edge's IETs.

Now we consider a node $v$ with $k$ neighbors.
From the viewpoint of node $v$, the sequence of events is a superposition of the events over its $k$ edges.
Because the $k$ nodes are statistically the same for $v$, it is sufficient to consider a $2(k+1)$-state Markov process with state space $\mathcal{S}_k = \{h, \ell\} \times \{h^k, h^{k-1}\ell, \ldots, h\ell^{k-1}, \ell^k\}$, where the first set in the product of the two sets represents the state of $v$, and the second set represents the states of $v$'s neighbors.

First, if $v$ is in state $h$, which happens with probability $p_h^*$ in the equilibrium, the IET distribution for $v$ depends on the states of $v$'s neighbors. 
The probability of finding exactly $k_h$ neighbors in state $h$ is given by $\binom{k}{k_h} p_h^{*k_h} (1-p_h^*)^{k-k_h}$, where $\binom{k}{k_h}$ is the binomial coefficient.
In this situation, $v$ experiences events produced by a Poisson process at rate $k_h \lambda_h + (k-k_h) \lambda_{\ell}$ because the superposition of the $k$ independent Poisson processes on edges with rate $\lambda_h$ or $\lambda_{\ell}$ is a Poisson process with the summed rate.
Second, if $v$ is in state $\ell$, which happens with probability $1-p_h^*$, all edges produce events at rate $\lambda_{\ell}$.
In this situation, $v$ experiences events produced by a Poisson process at rate $k \lambda_{\ell}$.
Therefore, an event occurs when $v$ is in state $h$ and it has $k_h$ neighbors in state $h$ with probability $p_h^* \binom{k}{k_h} p_h^{*k_h} (1-p_h^*)^{k-k_h} [k_h \lambda_h + (k-k_h) \lambda_{\ell}]^2 e^{-[k_h \lambda_h + (k-k_h) \lambda_{\ell}] \tau}/\Omega_k$ and when $v$ is in state $\ell$ with probability $(1-p_h^*) (k \lambda_{\ell})^2 e^{-k \lambda_{\ell} \tau} / \Omega_k$, where $\Omega_k = k \lambda_h p_h^{*2} + k \lambda_{\ell} (1-p_h^{*2})$.

\begin{widetext}
By combining these contributions, we derive the PDF for IETs on a node with $k$ neighbors as
\begin{align}\label{eq:node_PDF}
	f_k(\tau) &= \frac{p_h^*}{\Omega_k} \sum_{k_h=0}^k \binom{k}{k_h} p_h^{*k_h} (1-p_h^*)^{k-k_h} [k_h \lambda_h + (k-k_h) \lambda_{\ell}]^2 e^{-[k_h \lambda_h + (k-k_h) \lambda_{\ell}] \tau} + \frac{1-p_h^*}{\Omega_k} (k \lambda_{\ell})^2 e^{-k \lambda_{\ell} \tau} \nonumber \\
	&= \frac{e^{-k \lambda_{\ell} \tau}}{\Omega_k} \Bigg\{ p_h^* \left[ 1 - p_h^* (1-e^{-(\lambda_h - \lambda_{\ell})\tau}) \right]^k \Bigg[ \frac{k(\lambda_h-\lambda_{\ell})p_h^* e^{-(\lambda_h-\lambda_{\ell})\tau} (\lambda_h-\lambda_{\ell}+2)}{1 - p_h^* (1 - e^{-(\lambda_h-\lambda_{\ell})\tau})} + \nonumber \\
	& \quad + \frac{k (k-1) (\lambda_h-\lambda_{\ell})^2 p_h^{*2} e^{-2(\lambda_h-\lambda_{\ell})\tau}}{\left[ 1 - p_h^* (1 - e^{-(\lambda_h-\lambda_{\ell})\tau}) \right]^2} + (k \lambda_{\ell})^2 \Bigg] + (1-p_h^*) (k \lambda_{\ell})^2 \Bigg\}.
\end{align}



The first two moments of the PDF are given by 
\begin{eqnarray} \label{eq:node_first_moment}
	\langle \tau \rangle_k \equiv \int_0^{\infty} \tau f_k(\tau) d \tau = \frac{\,1\,}{\,\,\Omega_k}
\end{eqnarray}
and
\begin{align} \label{eq:node_second_moment}
	&\langle \tau^2 \rangle_k \equiv \int_0^{\infty} \tau^2 f_k(\tau) d \tau = \nonumber \\
                    &= \frac{2 p_h^*}{\Omega_k} \sum_{k_h=0}^k \binom{k}{k_h} \frac{p_h^{*k_h} (1-p_h^*)^{k-k_h}}{k_h \lambda_h + (k-k_h) \lambda_{\ell}} + \frac{2 (1-p_h^*)}{k \lambda_{\ell} \Omega_k}.
\end{align}
By substituting Eqs.~\eqref{eq:node_first_moment} and \eqref{eq:node_second_moment} into Eq.~\eqref{eq:CV}, we obtain the CV for node $v$ as
\begin{eqnarray} \label{eq:node_CV}
	\mathrm{CV}_k = \sqrt{ 2 k \left[ p_h^{*2} + (1-p_h^{*2}) \gamma \right] \left[ p_h^* \sum_{k_h=0}^k \binom{k}{k_h} \frac{p_h^{*k_h} (1-p_h^*)^{k-k_h}}{k_h + (k-k_h)\gamma} + \frac{1-p_h^*}{k\gamma} \right] -1 }.
\end{eqnarray}
Equation \eqref{eq:node_CV} reduces to  Eq.~\eqref{eq:edge_CV} when $k=1$.
\end{widetext}

For any $k$, Eq.~\eqref{eq:node_CV} yields $\displaystyle \lim_{\gamma \rightarrow 0} \text{CV}_k \rightarrow \infty $ and $\displaystyle \lim_{\gamma \rightarrow 1} \text{CV}_k = 1$.
In addition, the derivative of Eq.\eqref{eq:node_CV} with respect to $\gamma$ is negative for any $0< \gamma <1$.
Therefore, $\text{CV}_k$ monotonically decreases towards 1 as $\gamma \to 1$, which is consistent with Fig.~\ref{fig:CVheatmap_orig}.
Next, equating the derivative of the right-hand side of Eq.~\eqref{eq:edge_CV} with respect to $p_h^*$ to zero yields $p_h^* = 1/\sqrt{2} \approx 0.71$ for any $\gamma$.
The value of $p_h^*$ at which the derivative of the right-hand side of Eq.~\eqref{eq:node_CV} with respect to $p_h^*$ is equal to zero depends on $k$ and $\gamma$, but we numerically obtain $p_h^* \approx 0.7$ regardless of $k$ and $\gamma$.
Therefore, the $\text{CV}_k$ is large when $\gamma$ small and $p_h^* \approx 0.7$, which is consistent with Fig.~\ref{fig:CVheatmap_orig}.

\begin{figure*}[ht]
	\includegraphics{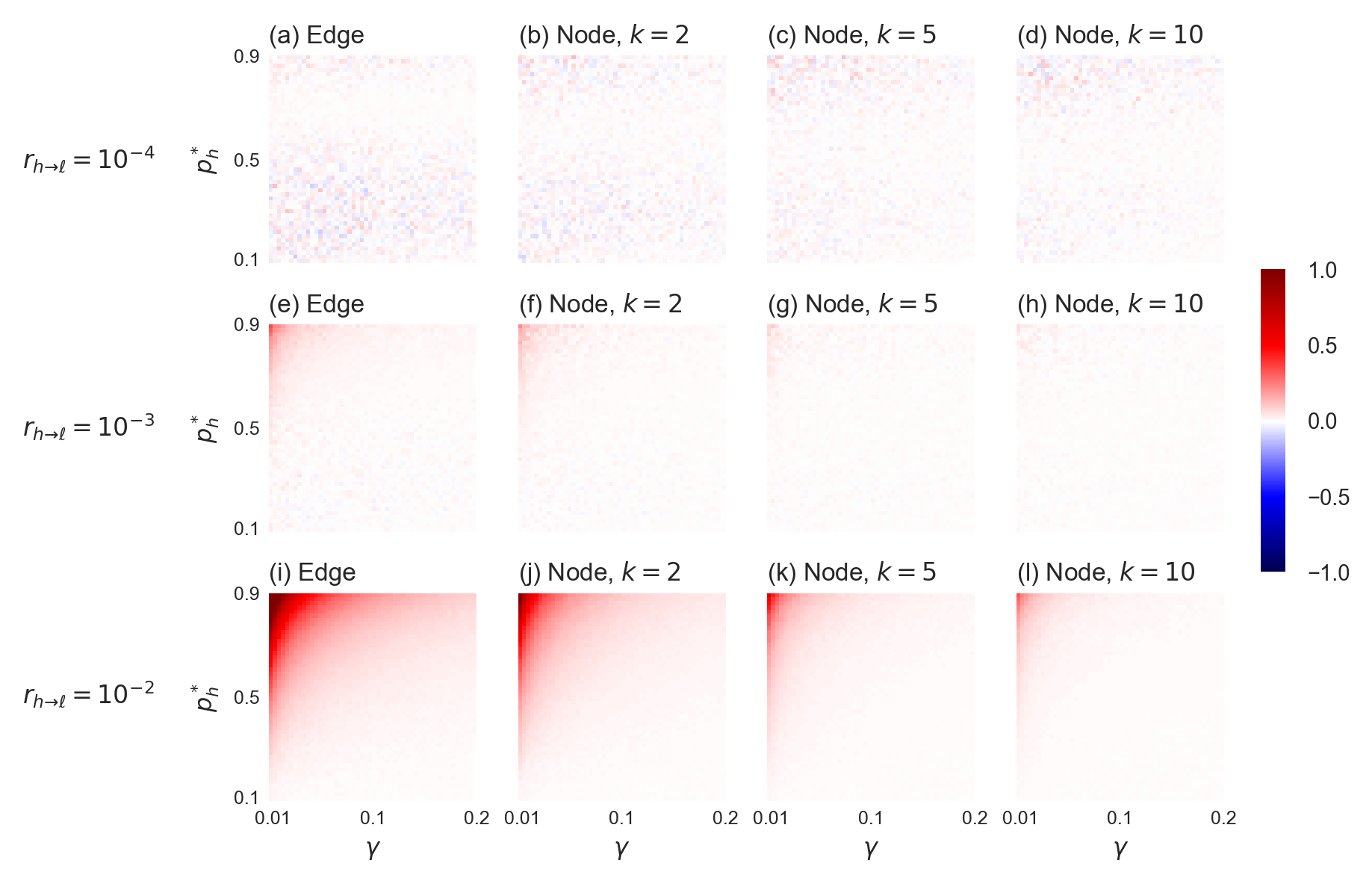}
	\caption{\label{fig:TSdifforig}
	Relative error between the analytically and numerically evaluated CV. 
	We set $r_{h \rightarrow \ell} = 10^{-4}$ in panels (a), (b), (c), and (d), $r_{h \rightarrow \ell} = 10^{-3}$ in panels (e), (f), (g), and (h), and $r_{h \rightarrow \ell} = 10^{-2}$ in panels (i), (j), (k), and ($\ell$).
	The results are for an edge (panels (a), (e), and (i)), a node with $k=2$ neighbors (panels (b), (f), and (j)), a node with $k=5$ neighbors (panels (c), (g), and (k)), and a node with $k=10$ neighbors (panels (d), (h), and ($\ell$)).
	}
\end{figure*}

To assess the accuracy of the theory, we calculated the relative error defined by [(theoretical CV) - (numerical CV)]/(numerical CV). 
Figures \ref{fig:TSdifforig}(a)--\ref{fig:TSdifforig}(d) show the relative error when $r_{h \rightarrow \ell} = 10^{-4}$.
In this case, the relative error is small across the entirety of our parameter region.
When $r_{h \rightarrow \ell}=10^{-3}$ (see Figs. \ref{fig:TSdifforig}(e)--\ref{fig:TSdifforig}(h)), and $r_{h \rightarrow \ell}=10^{-4}$ (see Figs. \ref{fig:TSdifforig}(i)--\ref{fig:TSdifforig}($\ell$)), the relative error is large when $\gamma$ is small and $p_h^*$ is large.

The relative error increases as $r_{h \rightarrow \ell}$ increases for the following reason.
A large value of $r_{h \rightarrow \ell}$ and moderate value of $p_h^*$ (i.e., $p_h^*$ that is not too close to 0 or 1) implies large $r_{\ell \rightarrow h}$, because $r_{\ell \rightarrow h} = r_{h \rightarrow \ell} p_h^*/(1-p_h^*)$.
When $r_{\ell \rightarrow h}$ is large, the mean IET for edges that are produced at event rate $\lambda_{\ell}$, which is equal to $1/\lambda_{\ell}$, is longer than the typical duration of the low-activity state of the edge (i.e., either $h \ell$ or $\ell^2$), which is proportional to $1 / r_{\ell \rightarrow h}$.
Note that, the low-activity state of an edge finishes only when both of the two nodes have transited from state $\ell$ to $h$, which occurs at rate $r_{\ell \rightarrow h}$ for each node.
Therefore, an edge is populated with sufficiently many IETs produced in the low-activity state of the edge if and only if $1/r_{\ell \rightarrow h} \gg 1/\lambda_{\ell}$, which leads to $r_{\ell \rightarrow h} \ll \lambda_{\ell}$.
According to our parametrization, we obtain $r_{\ell \rightarrow h} = r_{h \rightarrow \ell} \, p_h^* / (1-p_h^*)$ and $\lambda_{\ell} = \gamma \lambda_h = \gamma$, because we set $\lambda_h=1$.
Therefore, an edge is populated with sufficiently many IETs produced in its low-activity state if $r_{h \rightarrow \ell} \ll (1-p_h^*) \gamma / p_h^*$.
For example, when $\gamma = 0.01$ and $p_h^*=0.9$, we need $r_{h \rightarrow \ell} \ll (1-p_h^*) \gamma / p_h^* \approx 10^{-3}$.
This condition is violated when $r_{h \rightarrow \ell} = 10^{-3}$ or $10^{-2}$.
For these $r_{h \rightarrow \ell}$ values, the probability that events occur in the low-activity state of the edge is small.
However, our analytical derivation of the CV assumes that sufficiently many events and hence IETs occur in the typical duration of both low-activity and high-activity states of the edge at respective rates, i.e., $\lambda_{\ell}$ and $\lambda_h$.
This explains the discrepancy between the theoretical and numerical results observed in Figs.~\ref{fig:TSdifforig}(e)--($\ell$).

However, the model is still capable of producing large CV values even if few IETs are produced in the low-activity state of the edge.
In this situation, the IET between the last event of a high-activity period of the edge and the first event of the next high-activity period would generate a long IET, contributing to a relatively heterogeneous distribution of IETs.
This regime is not predicted by our analytical solution.

If $r_{h \rightarrow \ell} \approx \lambda_h$, then few IETs are produced during the $h^2$ state of the edge on average.
Therefore, a large CV value requires $r_{h \rightarrow \ell} \ll \lambda_h$, which is satisfied in Figs.~\ref{fig:CVheatmap_orig} and \ref{fig:TSdifforig} because the largest value of $r_{h \rightarrow \ell}$ that we use is $10^{-2}$ and we have set $\lambda_h = 1$.

\subsection{\label{sec:memory_coefficient}Correlation between consecutive interevent times}

In human activities, IETs for both edges and nodes are often positively correlated, i.e., long IETs tend to be followed by long IETs and vice versa \cite{Goh2008EPL, Karsai2012SciRep, HolmeSaramaki2012PhysRep, Karsai2018book}.
To examine this property, we compute the memory coefficient, $M$, of a sequence of IETs \cite{Goh2008EPL}, defined as
\begin{equation}
M \equiv \frac{1}{n-1} \sum_{i=1}^{n-1} \frac{(\tau_i-m_1)(\tau_{i+1}-m_2)}{\sigma_1 \sigma_2},
\end{equation}
where $n$ is the number of IETs in the sequence, $m_1$ and $\sigma_1$ are the average and standard deviation of $\{\tau_1, \tau_2, \ldots, \tau_{n-1} \}$, respectively, and $m_2$ and $\sigma_2$ are the average and standard deviation of $\{ \tau_2, \tau_3, \ldots, \tau_n \}$, respectively.
The memory coefficient measures the correlation coefficient of consecutive IETs, $(\tau_i, \tau_{i+1})$.

\begin{figure}[ht]
	\centering
	\includegraphics{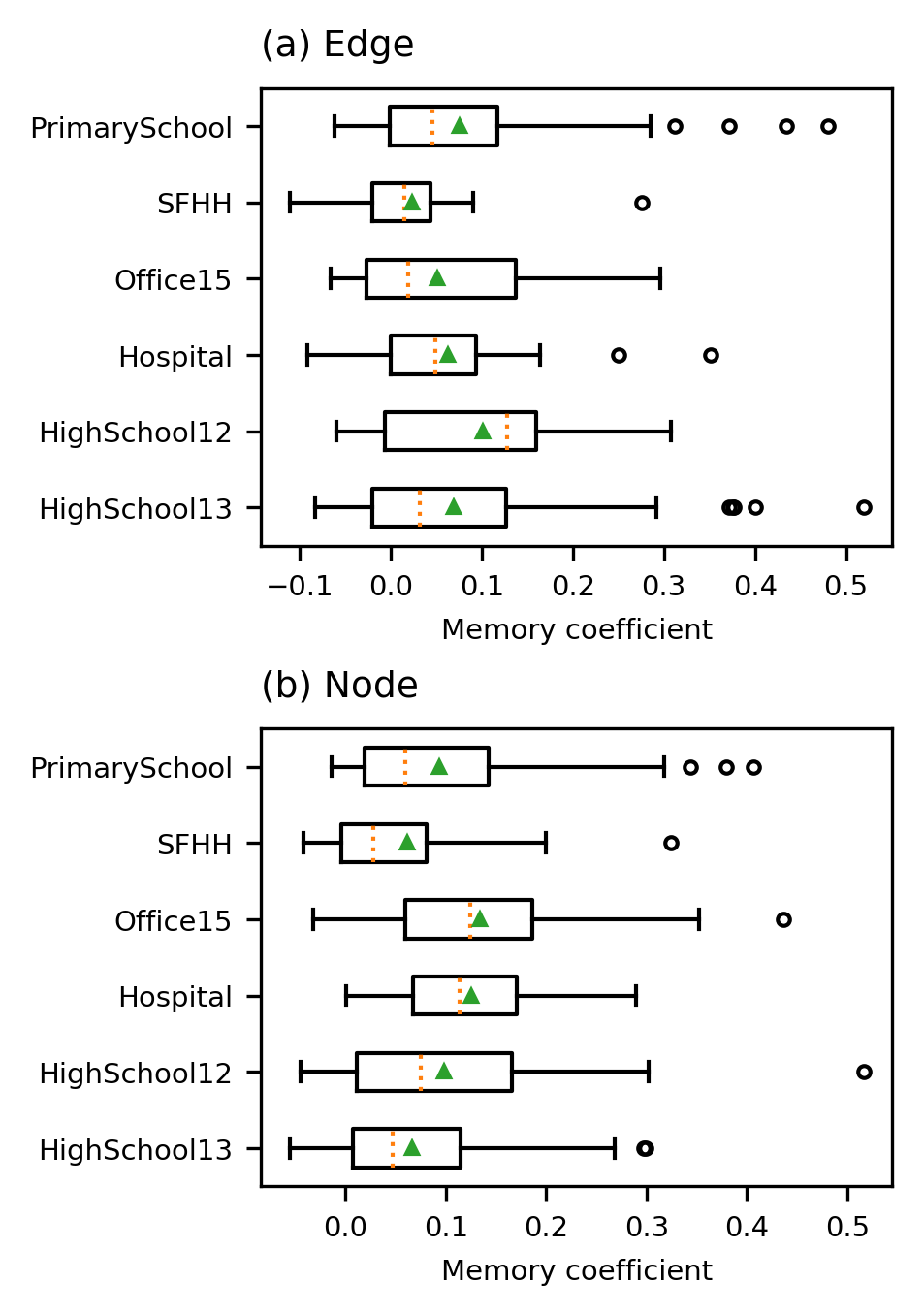}
	\caption{\label{fig:data_mem_coeff}
	Box plots of the memory coefficient for (a) edges and (b) nodes for the empirical data sets.
	The box shows the median (dotted line), the first quartile ($Q_1$), and the third quartile ($Q_3$); the whiskers show the minimum ($Q_1 - 1.5 \times \text{IQR}$), where IQR is the interquartile range, and the maximum ($Q_3 + 1.5 \times \text{IQR}$) values excluding outliers, where $\text{IQR} = Q_3 - Q_1$.
	The open circles are the outliers.
	The triangles are the sample means.
	}
\end{figure}

Figure \ref{fig:data_mem_coeff} shows the memory coefficient for the empirical data.
In all data sets, the edges show predominantly positive memory coefficients with values lying mostly between 0 and 0.1 (see Fig.~\ref{fig:data_mem_coeff}(a)).
The memory coefficient for nodes is also predominantly positive and tends to be larger than that for the edges (see Fig.~\ref{fig:data_mem_coeff}(b)).
These values are in accordance with previous results for various human activities \cite{Goh2008EPL}.

\begin{figure*}[ht]
	\includegraphics{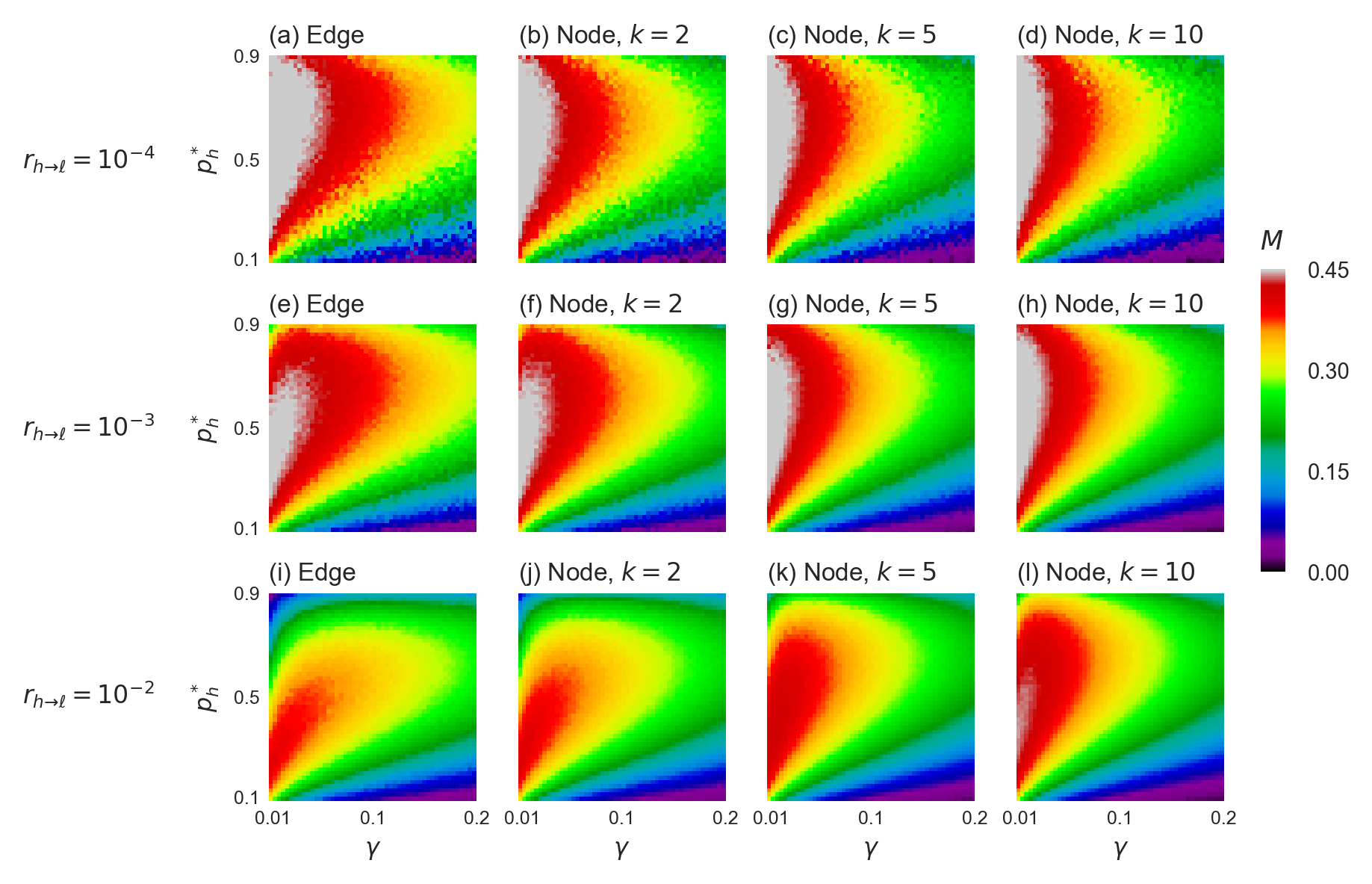}
	\caption{\label{fig:Mheatmap_orig}
	Memory coefficient for IETs generated by our original model.
	We set $r_{h \rightarrow \ell} = 10^{-4}$ in panels (a), (b), (c), and (d), $r_{h \rightarrow \ell} = 10^{-3}$ in panels (e), (f), (g), and (h), and $r_{h \rightarrow \ell} = 10^{-2}$ in panels (i), (j), (k), and ($\ell$).
	The results are for an edge (panels (a), (e), and (i)), a node with $k=2$ neighbors (panels (b), (f), and (j)), a node with $k=5$ neighbors (panels (c), (g), and (k)), and a node with $k=10$ neighbors (panels (d), (h), and ($\ell$)).
	For each set of parameter values, we generated IETs on the edges until all edges had at least $10^6$ events.
	}	
\end{figure*}

The memory coefficient for sequences of IETs generated by our model is shown in Fig.~\ref{fig:Mheatmap_orig}.
The figure indicates that the model produces positive $M$ for both edges and nodes, which is qualitatively consistent with the empirical data.
However, the memory coefficient values produced by the model are substantially larger than the empirical values.

\subsection{\label{sec:model_variants}Variants of the model}

In our original model, events occur on the edge at a higher rate if and only if both nodes forming the edge are in state $h$.
In this section, we study two variants of the model.
In the first variant, we assume that event on an edge occur at the higher rate, $\lambda_h$, if either node connected to the edge, not necessarily both nodes, is in state $h$.
Events on the edge occur at the lower rate $\lambda_{\ell}$ if and only if both nodes are in state $\ell$.
We call this variant the OR model.
An interpretation of the OR model is that, if an individual wants to interact with a neighbor, he/she can do so at the higher event rate regardless of whether or not the neighbor wants to interact.

\begin{figure*}[ht]
	\includegraphics{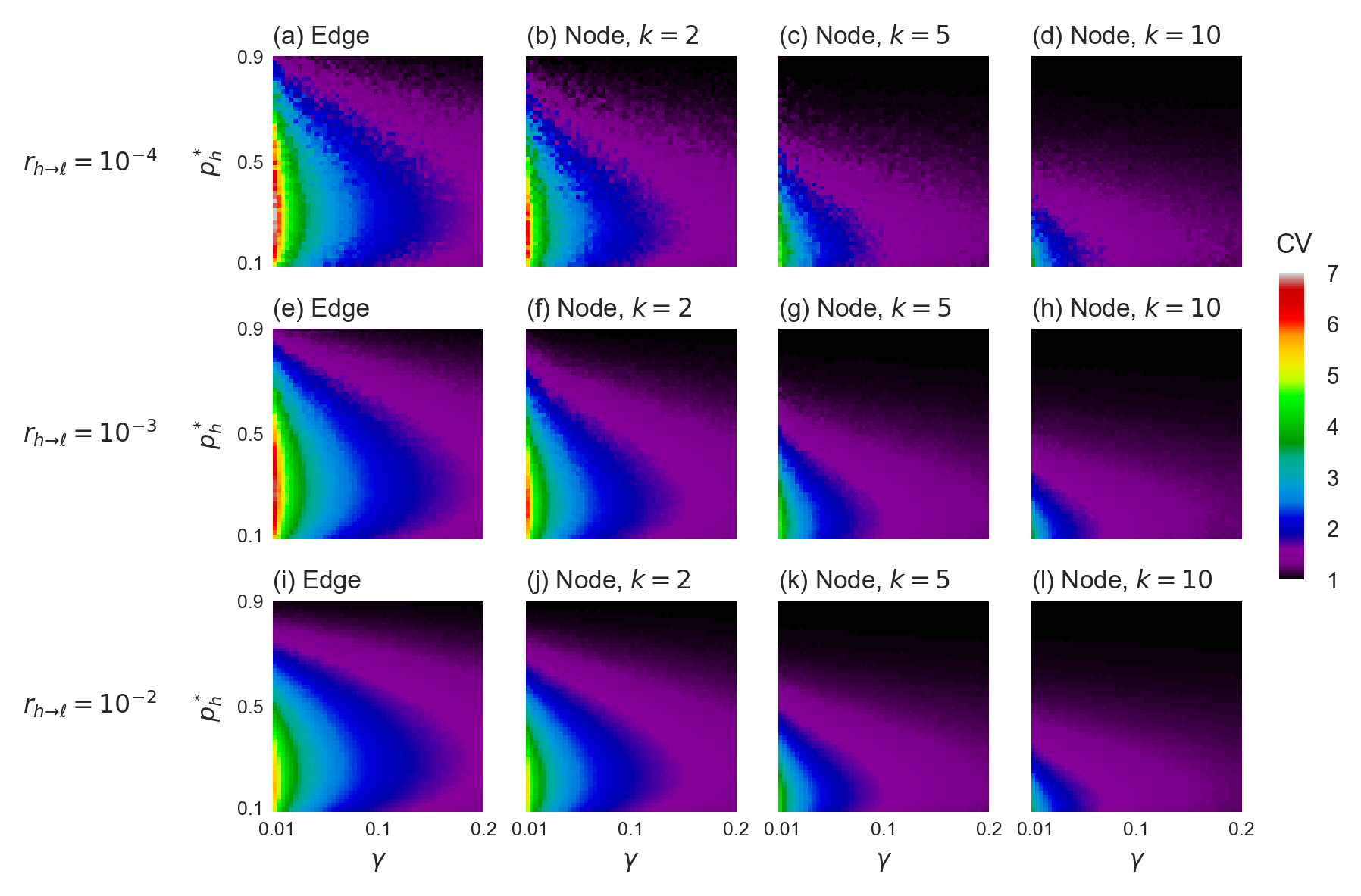}
	\caption{\label{fig:CVheatmap_OR}
	CV values for IETs generated by the OR model.
	We set $r_{h \rightarrow \ell} = 10^{-4}$ in panels (a), (b), (c), and (d), $r_{h \rightarrow \ell} = 10^{-3}$ in panels (e), (f), (g), and (h), and $r_{h \rightarrow \ell} = 10^{-2}$ in panels (i), (j), (k), and ($\ell$).
	The results are for an edge (panels (a), (e), and (i)), a node with $k=2$ neighbors (panels (b), (f), and (j)), a node with $k=5$ neighbors (panels (c), (g), and (k)), and a node with $k=10$ neighbors (panels (d), (h), and ($\ell$)).
	For each set of parameter values, we generated IETs on the edges until all edges had at least $10^6$ events.
	}	
\end{figure*}

For the OR model, we calculated the CV for IETs on edges and nodes by scanning the same values of $p_h^*$, $\gamma$, $r_{h \rightarrow \ell}$, and $k$ as those used in Fig.~\ref{fig:CVheatmap_orig}.
The results are shown in Fig.~\ref{fig:CVheatmap_OR}.
As in the original model, the OR model produces large CV values for both edges and nodes when $\gamma$ is small.
With respect to $p_h^*$, the OR model produces large CV values when $p_h^* \approx 0.3$ for the edge and that $p_h^*$ value that maximizes the CV decreases as $k$ increases.
This behavior is consistent with the analytical prediction (Appendix \ref{appendix:OR_CV}).
Unlike the original model, the region that the OR model produces large CV values for the node shrinks as $k$ increases.

In the second variant of the model, we assume that the node's $h$ and $\ell$ states independently contribute $\lambda_h$ and $\lambda_{\ell}$, respectively, to the event rate of the edge.
In other words, events occur on the edge at rate $2 \lambda_h$ if both nodes are in the $h$ state, $2 \lambda_{\ell}$ if both nodes are in the $\ell$ state, and $\lambda_h + \lambda_{\ell}$ if one node is in the $h$ state and the other node is in the $\ell$ state.
An interpretation of this variant of the model, which we call the IND model (named after ``independent''), is that the state of each individual independently contributes to the frequency of events between two individuals.

\begin{figure*}[ht]
	\includegraphics{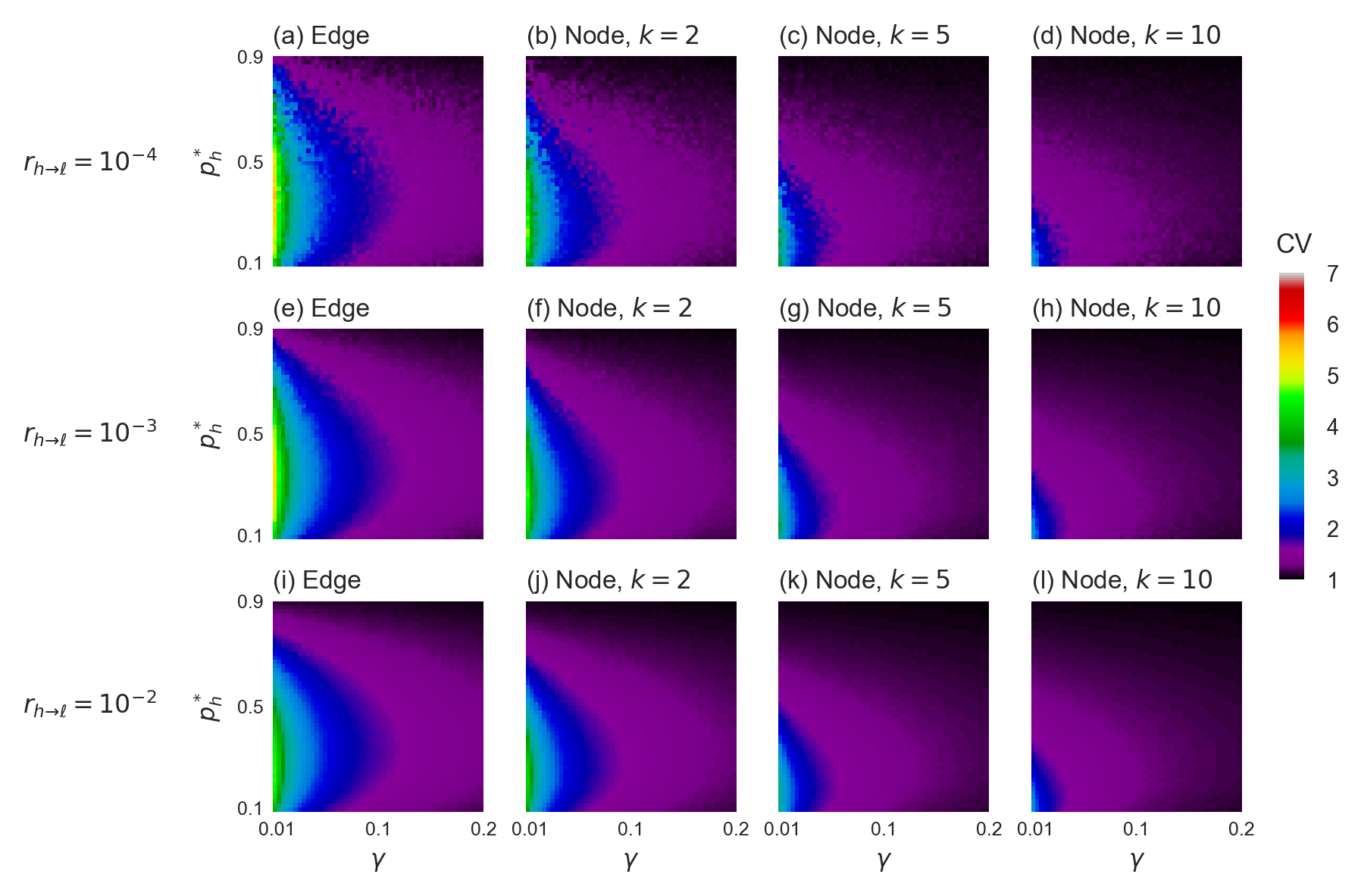}
	\caption{\label{fig:CVheatmap_IND}
	CV values for IETs generated by the IND model.
	We set $r_{h \rightarrow \ell} = 10^{-4}$ in panels (a), (b), (c), and (d), $r_{h \rightarrow \ell} = 10^{-3}$ in panels (e), (f), (g), and (h), and $r_{h \rightarrow \ell} = 10^{-2}$ in panels (i), (j), (k), and ($\ell$).
	The results are for an edge (panels (a), (e), and (i)), a node with $k=2$ neighbors (panels (b), (f), and (j)), a node with $k=5$ neighbors (panels (c), (g), and (k)), and a node with $k=10$ neighbors (panels (d), (h), and ($\ell$)).
	For each set of parameter values, we generated IETs on the edges until all edges had at least $10^6$ events.}
\end{figure*}

The CV values for the IND model are shown in Fig.~\ref{fig:CVheatmap_IND}.
Similarly to the original and the OR models, the IND model produces large CV values when $\gamma$ is small.
For any given $\gamma$, the IND model produces large values of CV when $0.3 \lessapprox p_h^* \lessapprox 0.4$ for the edge.
For the node, the $p_h^*$ value that maximizes the CV decreases as $k$ increases.
This behavior is consistent with the analytical result shown in Appendix \ref{appendix:IND_CV}.
The IND model behaves similarly to the OR model, i.e., it produces large CV values when $\gamma$ is small and $p_h^* \approx 0.3$, and the parameter region in which the node's CV is large  shrinks as $k$ increases.

\begin{figure}[ht]
	\includegraphics{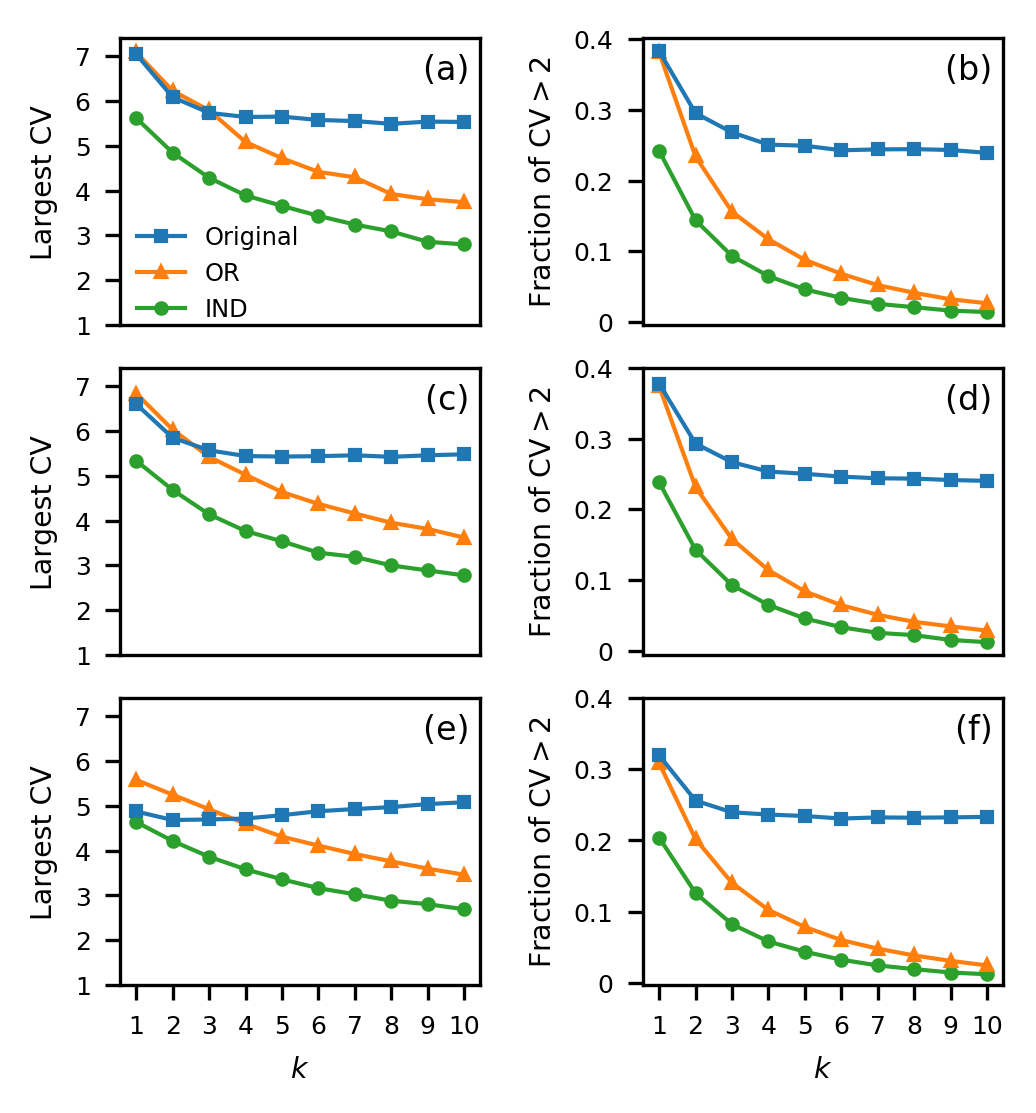}
	\caption{\label{fig:model_comparison}
	Comparison of CV values obtained from the original, OR, and IND models.
	Panels (a), (c), and (e) show the largest CV value in the ($\gamma$, $p_h^*$) parameter region explored in Figs.~\ref{fig:CVheatmap_orig}--\ref{fig:CVheatmap_IND}.
	Panels (b), (d), and (f) show the fraction of the ($\gamma$, $p_h^*$) pairs for which the CV values is larger than 2.
	Because $k$ represents the degree of the node, $k=1$ corresponds to the case of the single edge.
	We set $r_{h \rightarrow \ell}=10^{-4}$ in (a) and (b), $r_{h \rightarrow \ell}=10^{-3}$ in (c) and (d), and $r_{h \rightarrow \ell}=10^{-2}$ in (e) and (f).
	}
\end{figure}

To quantitatively compare the three models, we calculated two quantities.
First, we compute the largest CV value produced by each model when we vary $\gamma$ and $p_h^*$ in the parameter region used in Figs.~\ref{fig:CVheatmap_orig}, \ref{fig:CVheatmap_OR}, and \ref{fig:CVheatmap_IND}.
The largest CV value is compared among the three models in Fig.~\ref{fig:model_comparison}(a), where we set $r_{h \rightarrow \ell}=10^{-4}$ and vary the node's degree $k$.
The figure indicates that the original model consistently produces the largest CV values as $k$ increases, although the OR model produces comparably large CV values up to $k=3$.
Unlike the original model, the largest CV value produced by the OR and IND models visibly decreases as $k$ increases.
Second, we compute the fraction of the ($\gamma$, $p_h^*$) pairs for which the CV value is larger than two; a CV value larger than two is consistent with the results for empirical data (see Section \ref{sec:data_analysis}).
A large fraction value implies that a CV value larger than two is robustly produced for various parameter combinations.
The result for this analysis is shown in Fig.~\ref{fig:model_comparison}(b), where we again set $r_{h \rightarrow \ell}=10^{-4}$ and vary $k$.
The figure indicates that the original model has a larger fraction of the parameter region with CV larger than two than the OR and IND models.
The fraction remains roughly constant as $k$ increases for the original model, whereas it rapidly decreases as $k$ increases for the OR and the IND models. 
Figures \ref{fig:model_comparison}(a) and \ref{fig:model_comparison}(b) altogether suggest that the original model is more capable of producing large CVs of IETs on both edges and nodes than the OR and IND models, particularly when the node has a large degree.
The results are qualitatively the same for larger $r_{h \rightarrow \ell}$ values, i.e., $r_{h \rightarrow \ell}=10^{-3}$ (Figs.~\ref{fig:model_comparison}(c) and (d)) and $r_{h \rightarrow \ell}=10^{-2}$ (Figs.~\ref{fig:model_comparison}(e) and (f)).

\section{\label{sec:discussion}Discussion} 

We have started from the observation that heavy-tailed distributions of IETs are simultaneously present for both individual nodes and edges in the same empirical data.
We have proposed a continuous-time model and its variants for generating discrete events on edges that replicate this behavior to different extents.
Our main model crucially assumes that each node alternates between high-activity and low-activity states in a Markovian manner.
We showed that the original model, which requires that both nodes are in the high-activity state for the edge to have frequent events, is capable of producing large CV values for both individual nodes and edges in a broad parameter region.
The other two variants of the model are also capable of producing reasonably large CV values to some extent.
The proposed models allow interpretations.
For example, in the original model, two nodes are likely to interact if and only if both of them feel like interacting with others.

We have derived analytical solutions for our models by discarding the effects of IETs that contain state transitions of the edge.
The analytical solution was accurate when two conditions were met.
First, the ratio of the high- to low-activity event rates (i.e., $1/\gamma$) should not be extremely large, such as 100.
This condition is probably not unrealistic.
Second, the state transition rate of the node should be sufficiently small compared to the event rates on edges.
Under this condition, an epoch of the high- or low-activity state of an edge is typically long enough to host sufficiently many events at the constant rate, which is either $\lambda_h$ or $\lambda_{\ell}$.
This is the situation that the theory in Section \ref{sec:theoretical_cv} assumes.
Otherwise, a large fraction of IETs contains transitions of the edge's state, which cause a systematic discrepancy of the analytical expression from the numerical results.

In our models, each node switches between two states.
A possible extension of this assumption is to the case of  more than two states for each node.
Then, depending on how such a model translates the nodes' states into the event rate, the distribution of IETs on edges may be approximately a mixture of more than two exponential distributions, which may resemble or actually produce heavy-tailed distributions \cite{Feldmann1998PerfEval, Raghavan2014IEEETransComputSocSys, Masuda2018SIAM, Okada2020RSOS}.
In fact, a mixture of a small number of exponential distributions, including the case of just two exponential distributions, is often sufficient for approximating many empirical heavy-tailed distributions of IETs \cite{Okada2020RSOS, Jiang2016JStatMech, Raghavan2014IEEETransComputSocSys, Feldmann1998PerfEval}.
Therefore, one should carefully assess trade-offs between the complexity of extended models and the explanatory power of the model that one gains by assuming more states for nodes.

The distribution of IETs affects how disease and information spread across contact networks \cite{Min2011PRE, KarsaiPRE2011, Rocha2011PlosComputBio, Miritello2011PRE, Masuda2013F1000prime, Jo2014PRX, PastorSatorras2015RMF, Masuda2017book}.
Because many time-stamped event data probably have heavy-tailed distributions of IETs for both individual nodes and edges, the temporal network models proposed in the present study are expected to be useful for modeling dynamical processes on temporal networks including contagion processes.
It seems that model-based studies of contagion processes on temporal networks have not paid much attention to the simultaneous presence of heavy-tailed distributions of IETs on nodes and edges \cite{Karsai2018book}.
How this property affects key indicators of contagion processes such as the epidemic threshold, the final epidemic size, and equilibrium fraction of infected nodes, as well as indicators of other dynamical processes, warrants future work.

\begin{acknowledgments}
We thank the SocioPatterns collaboration (http://www.sociopatterns.org) for providing the data sets.
EFdR thanks the financial support by the Coordena\c{c}\~{a}o de Aperfei\c{c}oamento de Pessoal de N\'{i}vel Superior - Brasil (CAPES) - Finance Code 001.
AL acknowledges the Cross-Disciplinary Fellowship Award (Grant: LT000696/2018-C) from International Human Frontier Science Program and Foster Lab at University of Oxford.
NM thanks the financial support by AFOSR European Office (under Grant no. FA9550-19-1-7024.
\end{acknowledgments}

\appendix

\section{Distributions of interevent times for the other data sets} \label{appendix:data_survival}

\begin{figure}[ht]
	\includegraphics{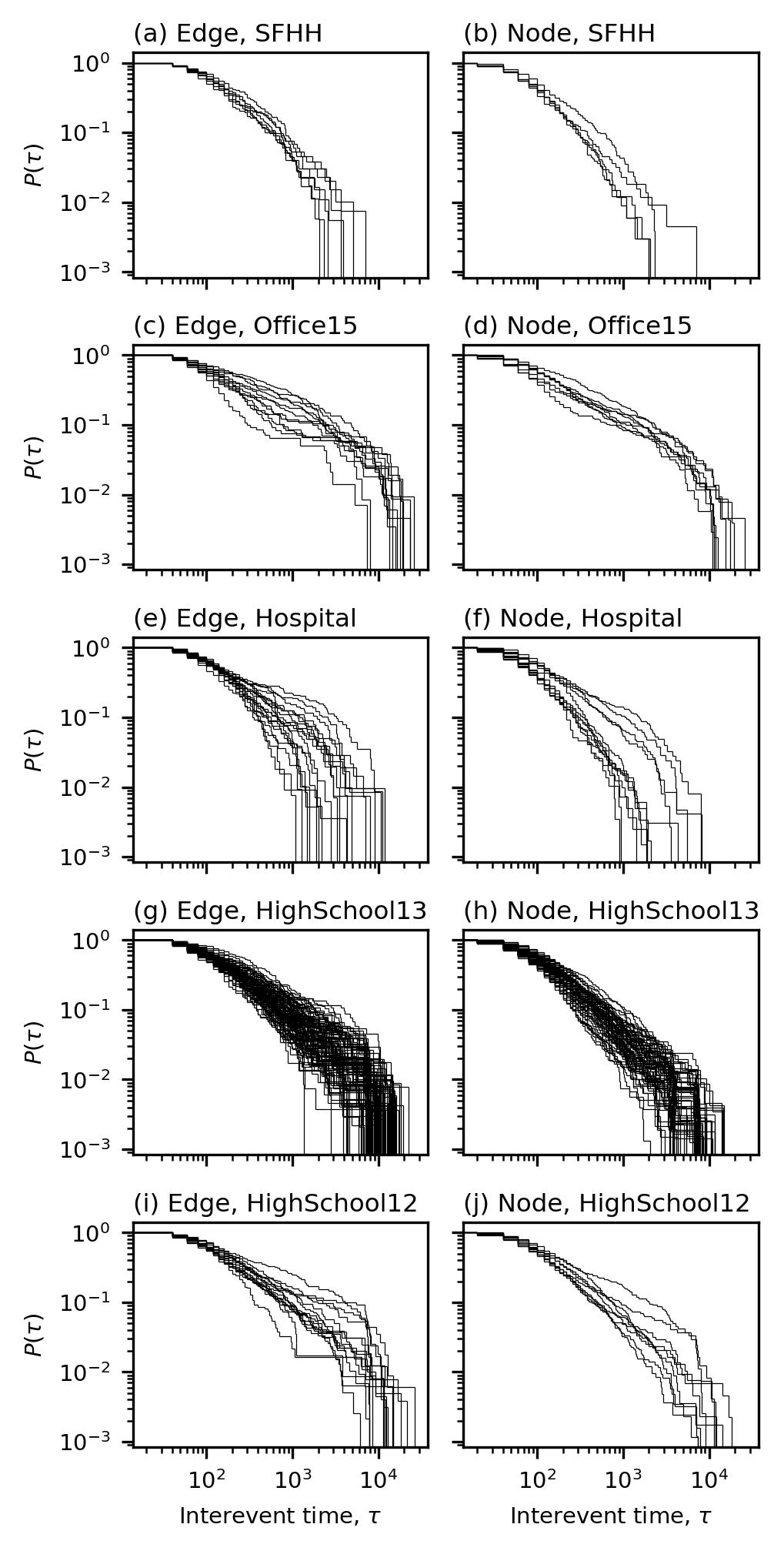}
	\caption{\label{fig:data_survival}
		Survival function, $P(\tau)$, of IETs on edges and nodes for the different data sets.
	}
\end{figure}

Figure \ref{fig:data_survival} shows the survival function of IETs for the different data sets.

\section{Analytical evaluation of the CV of interevent times for the OR model} \label{appendix:OR_CV}

In this section, we analytically examine the CV of IETs for the OR model.
In the OR model, events occur at the lower rate $\lambda_{\ell}$ if and only if the two nodes forming an edge are in state $\ell$.
Therefore, using the same reasoning as that for the original model in Sec.~\ref{sec:theoretical_cv}, one obtains the PDF of IETs for an edge as follows:
\begin{eqnarray}\label{eq:edge_PDF_OR}
	f_{\rm edge}^{\rm \scriptscriptstyle OR}(\tau) = \frac{\lambda_{\ell} p_{\ell}^*}{\Omega_1^{\rm \scriptscriptstyle OR}} \lambda_{\ell} e^{- \lambda_{\ell} \tau} + \frac{\lambda_h (1 - p_{\ell}^*)}{\Omega_1^{\rm \scriptscriptstyle OR}} \lambda_h e^{- \lambda_h \tau},
\end{eqnarray}
where $\Omega_1^{\rm \scriptscriptstyle OR} = \lambda_{\ell} p_{\ell}^* + \lambda_h (1 - p_{\ell}^{*2})$ and, for convenience, we have used $p_{\ell}^* = 1 - p_h^*$.

The first two moments of this PDF are given by
\begin{eqnarray} \label{eq:edge_first_moment_OR}
	\langle \tau \rangle_{\text{edge}} \equiv \int_0^\infty \tau f_{\text{edge}}^{\rm \scriptscriptstyle OR} (\tau) d \tau = \frac{\,1\,}{\,\,\Omega_1^{\rm \scriptscriptstyle OR}}
\end{eqnarray}
and
\begin{equation} \label{eq:edge_second_moment_OR}
	\langle \tau^2 \rangle_{\text{edge}} \equiv \int_0^\infty \tau^2 f_{\text{edge}}^{\rm \scriptscriptstyle OR}(\tau) d \tau  = \frac{\,2\,}{\,\,\Omega_1^{\rm \scriptscriptstyle OR}}\left[ \frac{p_{\ell}^{*2}}{\lambda_{\ell}} + \frac{(1-p_{\ell}^{*2})}{\lambda_h} \right].
\end{equation}

By substituting Eqs.~\eqref{eq:edge_first_moment_OR} and \eqref{eq:edge_second_moment_OR} into Eq.~\eqref{eq:CV}, we obtain
\begin{eqnarray} \label{eq:edge_CV_OR}
	\mathrm{CV}_{\text{edge}}^{\rm \scriptscriptstyle OR} = \sqrt{ 1 +  \frac{2 p_{\ell}^{*2} (1-p_{\ell}^{*2}) (1-\gamma)^2}{\gamma} },
\end{eqnarray}
where $\mathrm{CV}_{\text{edge}}^{\rm \scriptscriptstyle OR}$ is the CV for the edge's IETs for the OR model.

We proceed with the same steps as those in Sec.~\ref{sec:theoretical_cv} to calculate the CV for a node with $k$ neighbors as follows.
The PDF for a node with $k$ neighbors is given by
\begin{align} \label{eq:node_PDF_OR}
	f_k^{\rm \scriptscriptstyle OR}(\tau) =& \frac{p_{\ell}^*}{\Omega_k^{\rm \scriptscriptstyle OR}} \sum_{k_{\ell}=0}^k \binom{k}{k_{\ell}}  p_{\ell}^{*k_{\ell}} (1-p_{\ell}^*)^{k-k_{\ell}} [ \lambda_{\ell} k_{\ell} \nonumber \\
         &+ \lambda_h (k - k_{\ell}) ]^2 e^{-\left[ \lambda_{\ell} k_{\ell} + \lambda_h (k - k_{\ell}) \right]\tau} \nonumber \\
         &+ \frac{1 - p_{\ell}^*}{\Omega_k^{\rm \scriptscriptstyle OR}} (k \lambda_h)^2 e^{-k \lambda_h \tau},
\end{align}
where $k_{\ell}$ is the number of neighbors in the $\ell$ state, and $\Omega_k^{\rm \scriptscriptstyle OR} = k \lambda_{\ell} p_{\ell}^{*2} + k \lambda_h (1 - p_{\ell}^{*2})$.

The first two moments of $f_k^{\rm \scriptscriptstyle OR}(\tau)$ are given by
\begin{eqnarray} \label{eq:node_first_moment_OR}
	\langle \tau \rangle_k \equiv \int_0^{\infty} \tau f_k(\tau) d \tau = \frac{\,1\,}{\,\,\Omega_k^{\rm \scriptscriptstyle OR}}
\end{eqnarray}
and
\begin{align} \label{eq:node_second_moment_OR}
	&\langle \tau^2 \rangle_k \equiv \int_0^{\infty} \tau^2 f_k(\tau) d \tau = \nonumber \\
                    &= \frac{2 p_{\ell}^*}{\Omega_k^{\rm \scriptscriptstyle OR}} \sum_{k_{\ell}=0}^k \binom{k}{k_{\ell}} \frac{p_{\ell}^{*k_{\ell}} (1-p_{\ell}^*)^{k-k_{\ell}}}{k_{\ell} \lambda_{\ell} + (k-k_{\ell}) \lambda_h} + \frac{2 (1-p_{\ell}^*)}{k \lambda_h \Omega_k^{\rm \scriptscriptstyle OR}}.
\end{align}
By substituting Eqs.~\eqref{eq:node_first_moment_OR} and \eqref{eq:node_second_moment_OR} into Eq.~\eqref{eq:CV}, we obtain the CV for node $v$ as
\begin{widetext}
\begin{eqnarray} \label{eq:node_CV_OR}
	\mathrm{CV}_k^{\rm \scriptscriptstyle OR} = \sqrt{ 2 k \left[ p_{\ell}^{*2} + (1-p_{\ell}^{*2}) \gamma \right] \left[ p_{\ell}^* \sum_{k_{\ell}=0}^k \binom{k}{k_{\ell}} \frac{p_{\ell}^{*k_{\ell}} (1-p_{\ell}^*)^{k-k_{\ell}}}{k_{\ell} + (k-k_{\ell})\gamma} + \frac{1-p_{\ell}^*}{k\gamma} \right] -1 },
\end{eqnarray}
\end{widetext}
which generalizes Eq.~\eqref{eq:edge_CV_OR}.

\begin{figure}[ht]
	\includegraphics{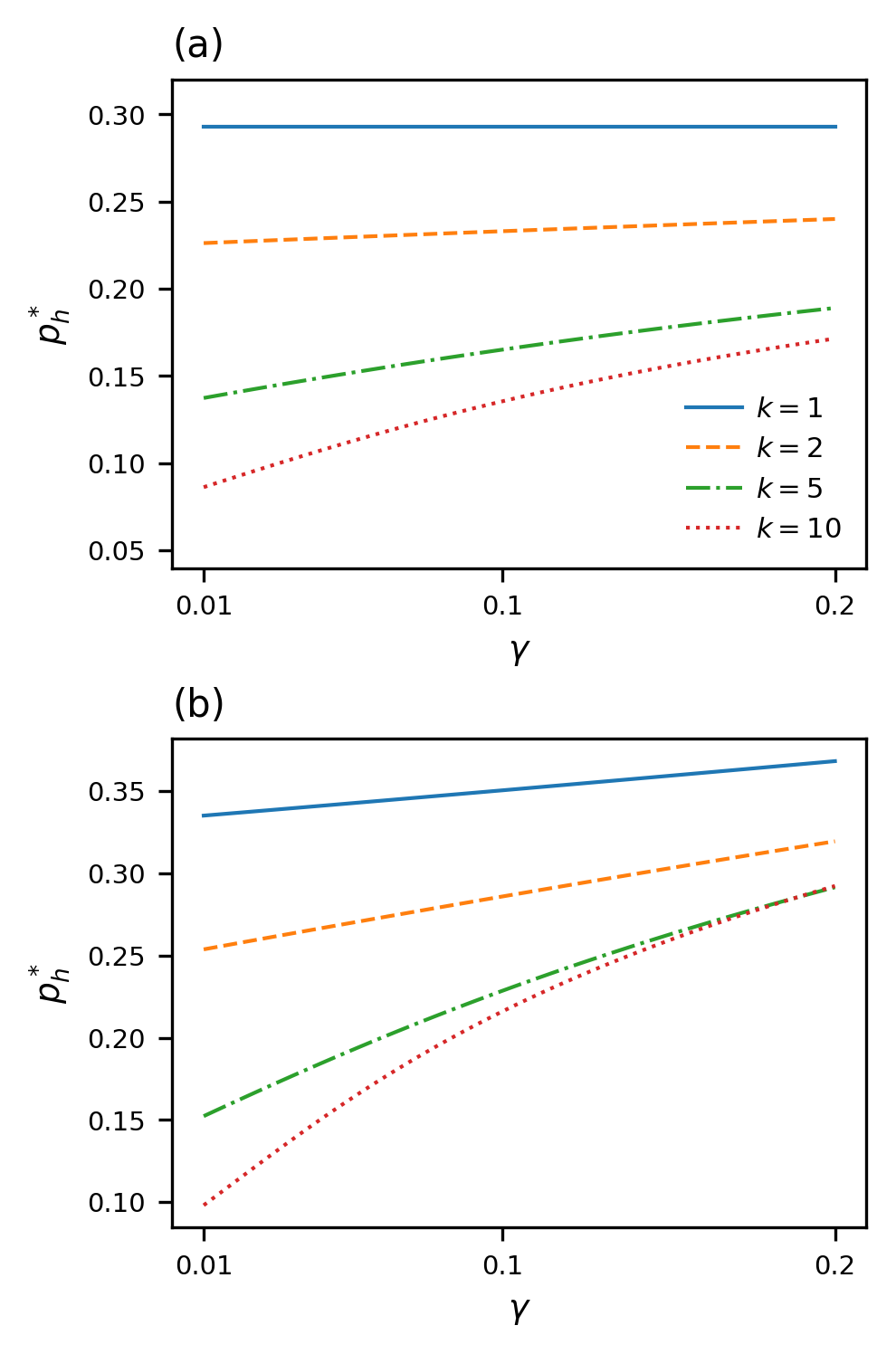}
	\caption{\label{fig:argmaxCV_OR_IND}
		Values of $p_h^*$ that yield the maximum of (a) $\text{CV}_k^{\rm \scriptscriptstyle OR}$ and (b) $\text{CV}_k^{\rm \scriptscriptstyle IND}$ for different values of $\gamma$ and $k$.
	}
\end{figure}

The behavior of Eq.~\eqref{eq:node_CV_OR} with respect to $\gamma$ is qualitatively the same as that of the original model (i.e., Eq.~\eqref{eq:node_CV}).
In other words, $\displaystyle \lim_{\gamma \to 0} \text{CV}_k^{\rm \scriptscriptstyle OR} \to \infty$, $\lim_{\gamma \to 1} \text{CV}_k^{\rm \scriptscriptstyle OR} = 1$, and $\text{CV}_k^{\rm \scriptscriptstyle OR}$ monotonically decreases as $\gamma$ increases, which is consistent with Fig.~\ref{fig:CVheatmap_OR}.
The extremum of Eq.~\eqref{eq:edge_CV_OR} with respect to $p_{\ell}^*$ occurs at $p_{\ell}^* = 1/\sqrt{2} \approx 0.71$, hence, $p_h^* \approx 0.29$, for any $\gamma$.
Therefore, the CV for the edge is large when $\gamma$ is small and $p_h^* \approx 0.3$.
By contrast, the value of $p_h^*$ that maximizes Eq.~\eqref{eq:node_CV_OR} for a given $\gamma$ value strongly depends on $k$.
In Fig.~\ref{fig:argmaxCV_OR_IND}(a), we numerically inspect this dependence.
The value of $p_h^*$ that maximizes Eq.~\eqref{eq:node_CV_OR} decreases as $k$ increases, and, for a given value of $k$, it monotonically increases as $\gamma$ increases.
These results are consistent with Fig.~\ref{fig:CVheatmap_OR}.

\hfill \break
\section{Analytical evaluation of the CV of interevent times for the IND model} \label{appendix:IND_CV}

In this section, we analytically examine the CV of IETs for the IND model.
In the IND model, events on an edge occur at rate $2 \lambda_h$ if both nodes are in state $h$, at rate $\lambda_h + \lambda_{\ell}$ if one node is the $h$ and the other node is in the $\ell$ state, and at rate $2 \lambda_{\ell}$ if both nodes are in the $\ell$ state.
Therefore, the PDF of IETs for an edge is given by
\begin{align}\label{eq:edge_PDF_IND}
	f_{\rm edge}^{\rm \scriptscriptstyle IND}(\tau) =& \frac{2 \lambda_h p_h^{*2}}{\Omega_1^{\rm \scriptscriptstyle IND}} 2 \lambda_h e^{-2 \lambda_h \tau} \nonumber \\
	& + \frac{2(\lambda_h+\lambda_{\ell})p_h^*(1-p_h^*)}{\Omega_1^{\rm \scriptscriptstyle IND}} (\lambda_h+\lambda_{\ell}) e^{-(\lambda_h+\lambda_{\ell})\tau} \nonumber \\
	& + \frac{2\lambda_{\ell}(1-p_h^*)^2}{\Omega_1^{\rm \scriptscriptstyle IND}} 2 \lambda_{\ell} e^{-2\lambda_{\ell}\tau},
\end{align}
where $\Omega_1^{\rm \scriptscriptstyle IND}=2\lambda_h p_h^{*2}+2(\lambda_h+\lambda_{\ell})p_h^*(1-p_h^*)+2\lambda_{\ell}p_h^{*2}$.

The first two moments of $f_{\rm edge}^{\rm \scriptscriptstyle IND}(\tau)$ are give by
\begin{eqnarray} \label{eq:edge_first_moment_IND}
	\langle \tau \rangle_{\rm edge} \equiv \int_0^{\infty} \tau f_{\rm edge}^{\rm \scriptscriptstyle IND} d \tau = \frac{\,1\,}{\,\,\Omega_1^{\rm \scriptscriptstyle IND}}
\end{eqnarray}
and
\begin{align} \label{eq:edge_second_moment_IND}
	&\langle \tau^2 \rangle_{\text{edge}} \equiv \int_0^\infty \tau^2 f_{\text{edge}}^{\rm OR}(\tau) d \tau \nonumber \\
	&  = \frac{\,2\,}{\,\,\Omega_1^{\rm \scriptscriptstyle IND}}\left[ \frac{p_h^{*2}}{2\lambda_h} + \frac{2p_h^*(1-p_h^*)}{\lambda_h+\lambda_{\ell}} + \frac{(1-p_h^*)^2}{2\lambda_{\ell}} \right].
\end{align}

\hfill \break
\hfill \break
By substituting Eqs.~\eqref{eq:edge_first_moment_IND} and \eqref{eq:edge_second_moment_IND} into Eq.~\eqref{eq:CV}, we obtain
\begin{eqnarray} \label{eq:edge_CV_IND}
	\mathrm{CV}_{\text{edge}}^{\rm \scriptscriptstyle IND} = \sqrt{ 1 +  \frac{2 p_h^* \left[ 1-(2-\gamma)p_h^* + (1-\gamma)p_h^{*2} \right]}{\gamma (1+\gamma)}},\quad
\end{eqnarray}
where $\mathrm{CV}_{\text{edge}}^{\rm \scriptscriptstyle IND}$ is the CV for the edge's IETs for the IND model.

The PDF for IETs on a node with $k$ neighbors is given by
\begin{align} \label{eq:node_PDF_IND}
	f_k^{\rm \scriptscriptstyle IND}(\tau) =& \sum_{k_h=0}^k \binom{k}{k_h} \frac{p_h^{*k_h}(1-p_h^*)^{k-k_h}}{\Omega_k^{\rm \scriptscriptstyle IND}} \{ p_h^* [ \lambda_h(k+k_h) \nonumber\\
        &+ \lambda_{\ell} (k-k_h) ]^2 e^{-\left[ \lambda_h(k+k_h) + \lambda_{\ell} (k-k_h) \right]\tau} + (1-p_h^*) \nonumber\\
        & \times [ \lambda_h k_h + \lambda_{\ell}(2k-k_h) ]^2 e^{-\left[ \lambda_h k_h + \lambda_{\ell}(2k-k_h) \right]\tau} \},
\end{align}
where $\Omega_k^{\rm \scriptscriptstyle IND} = 2 k \left[ \lambda_h p_h^* + \lambda_{\ell} (1-p_h^*) \right]$.

\begin{widetext}
The first two moments of $f_k^{\rm \scriptscriptstyle IND}(\tau)$ are given by
\begin{eqnarray} \label{eq:node_first_moment_IND}
	\langle \tau \rangle_k \equiv \int_0^{\infty} \tau f_k(\tau) d \tau = \frac{\,1\,}{\,\,\Omega_k^{\rm \scriptscriptstyle IND}}
\end{eqnarray}
and
\begin{align} \label{eq:node_second_moment_IND}
	&\langle \tau^2 \rangle_k \equiv \int_0^{\infty} \tau^2 f_k(\tau) d \tau = \nonumber \\
                    &= \sum_{k_h=0}^k \binom{k}{k_h} \frac{p_h^{*k_h}(1-p_h^*)^{k-k_h}}{\Omega_k^{\rm \scriptscriptstyle IND}} \Bigg\{ \frac{2p_h^*}{\lambda_h(k+k_h)+\lambda_{\ell}(k-k_h)} + \nonumber \\
			& + \frac{2(1-p_h^*)}{\lambda_h k_h + \lambda_{\ell}(2k-k_h)} \Bigg\}.
\end{align}

By substituting Eqs.~\eqref{eq:node_first_moment_IND} and \eqref{eq:node_second_moment_IND} into Eq.~\eqref{eq:CV}, we obtain the CV for node $v$ as
\begin{eqnarray} \label{eq:node_CV_IND}
	\mathrm{CV}_k^{\rm \scriptscriptstyle IND} = \sqrt{ 4k \left[ p_h^* + \gamma(1-p_h^*) \right] \sum_{k_h=0}^k \binom{k}{k_h} p_h^{*k_h}(1-p_h^*)^{k-k_h} \Bigg[ \frac{p_h^*}{k+k_h+\gamma(k-k_h)} + \frac{1-p_h^*}{k_h+\gamma(2k-k_h)} \Bigg] - 1 },
\end{eqnarray}
which generalizes Eq.~\eqref{eq:edge_CV_IND}.

The behavior of Eq.~\eqref{eq:node_CV_IND} with respect to $\gamma$ is qualitatively the same as that of the original model.
In other words, $\displaystyle \lim_{\gamma \to 0} \text{CV}_k^{\rm \scriptscriptstyle IND} \to \infty$, $\displaystyle \lim_{\gamma \to 1} \text{CV}_k^{\rm \scriptscriptstyle IND} = 1$, and the $\text{CV}_k^{\rm \scriptscriptstyle IND}$ monotonically decreases as $\gamma$ increases, which are consistent with Fig.~\ref{fig:CVheatmap_IND}.
The $p_h^*$ value that maximizes Eq.~\eqref{eq:edge_CV_IND} is given by 
\begin{eqnarray}
	p_h^* = \frac{2-\gamma-\sqrt{1-\gamma+\gamma^2}}{3(1-\gamma)},
\end{eqnarray}
which is plotted as the solid line in Fig.~\ref{fig:argmaxCV_OR_IND}(b).
The numerically obtained $p_h^*$ value that maximizes Eq.~\eqref{eq:node_CV_IND} decreases as $k$ increases, and increases as $\gamma$ increases (see Fig.~\ref{fig:argmaxCV_OR_IND}(b)).
These results are consistent with Fig.~\ref{fig:CVheatmap_IND}.
\end{widetext}



%

\end{document}